\documentclass[prd,twocolumn,showpacs,nofootinbib, aps]{revtex4-1}
\usepackage{bm}
\usepackage{graphicx}
\usepackage{multirow}
\usepackage{amsmath}
\usepackage{amssymb}
\usepackage{tabulary}
\usepackage{hyperref}
\usepackage{color}
\usepackage{xspace}
\usepackage{verbatim}

\newcommand{\vmin}{\ifmmode v_{\mathrm{min}} \else $v_{\mathrm{min}}$ \fi\xspace}
\newcommand{\vesc}{\ifmmode v_{\mathrm{esc}} \else $v_{\mathrm{esc}}$ \fi\xspace}
\newcommand{\sigmapsi}{\relax \ifmmode \sigma_{\mathrm{p}}^{\mathrm{SI}}\else $\sigma_{\mathrm{p}}^{\mathrm{SI}}$\fi\xspace}
\newcommand{\sigmaNsi}{\ifmmode \sigma_{\mathrm{N}}^{\mathrm{SI}} \else $\sigma_{\mathrm{N}}^{\mathrm{SI}}$\fi\xspace}
\newcommand{\sigmapsd}{\relax \ifmmode \sigma_{\mathrm{p}}^{\mathrm{SD}}\else $\sigma_{\mathrm{p}}^{\mathrm{SD}}$\fi\xspace}
\newcommand{\sigmansd}{\ifmmode \sigma_{\mathrm{n}}^{\mathrm{SD}}\else $\sigma_{\mathrm{n}}^{\mathrm{SD}}$\fi\xspace}

\newcommand{\kms}{\ifmmode \textrm{ km s}^{-1}\else $\textrm{ km s}^{-1}$\fi\xspace}
\newcommand{\qhat}{\hat{q}}

\allowdisplaybreaks

\newcommand{\dbd}[2]{\ifmmode \frac{\textrm{d}#1}{\textrm{d}#2}\else $\textrm{d}#1/\textrm{d}#2$\fi\xspace}

\newcommand{\Op}[1]{\mathcal{O}_{#1}}

\begin{document}

\hfill SACLAY-t15/093

\title{New directional signatures from the non-relativistic effective field theory of dark matter}
\author{Bradley J. Kavanagh}
\email{bradley.kavanagh@cea.fr}
\affiliation{Institut de physique th\'eorique, Universit\'e Paris Saclay, CNRS, CEA, F-91191 Gif-sur-Yvette, France}

\begin{abstract}
The framework of non-relativistic effective field theory (NREFT) aims to generalise the standard analysis of direct detection experiments in terms of spin-dependent (SD) and spin-independent (SI) interactions. We show that a number of NREFT operators lead to distinctive new directional signatures, such as prominent ring-like features in the directional recoil rate, even for relatively low mass WIMPs. We discuss these signatures and how they could affect the interpretation of future results from directional detectors. We demonstrate that considering a range of possible operators introduces a factor of 2 uncertainty in the number of events required to confirm the median recoil direction of the signal.  Furthermore, using directional detection, it is possible to distinguish the more general NREFT interactions from the standard SI/SD interactions at the $2\sigma$ level with $\mathcal{O}(100-500)$ events. In particular, we demonstrate that for certain NREFT operators, directional sensitivity provides the only method of distinguishing them from these standard operators, highlighting the importance of directional detectors in probing the particle physics of dark matter. 
\end{abstract}

\maketitle

\section{Introduction}
\label{sec:introduction}

The detection of dark matter (DM) in a laboratory setting is considered one of the greatest goals of modern particle physics. Direct detection experiments \cite{Goodman:1984dc, Drukier:1986tm} aim to measure the keV-scale recoil energy imparted on detector nuclei by interactions with Weakly Interacting Massive Particles (WIMPs) in the Galactic halo. The motion of the Earth and Sun in the Galactic rest frame induces a highly directional flow of DM particles in the lab frame, known as the `WIMP wind'. The result is that the mean recoil direction is opposite that of the Solar motion, in the direction of the constellation Vela. Detection of this directional signal would provide strong evidence for a DM origin of the signal \cite{Spergel:1987kx, Copi:1999pw, Copi:2002hm}. 

A number of experiments with directional sensitivity are currently in development. The most advanced of these are low pressure time projection chambers (TPCs) such as DRIFT \cite{Daw:2011wq, Battat:2014van}, MIMAC \cite{Riffard:2013psa, Riffard:2015rga}, DMTPC \cite{Monroe:2011er, Battat:2013gma}, NEWAGE \cite{Miuchi:2010hn,Miuchi:2011qw} and D3 \cite{Vahsen:2011qx, Vahsen:2015oya}, though a number of other technologies are also being considered, including nuclear emulsions \cite{Naka:2011sf}, DNA-based techniques \cite{Drukier:2012hj} and the possibility of exploiting columnar recombination in Xenon targets \cite{Nygren:2013nda, Mohlabeng:2015efa}. The analysis of data from such experiments (as well as from non-directional experiments) typically assumes standard spin-dependent (SD) or spin-independent (SI) contact interactions \cite{Jungman:1995df, Cerdeno:2010jj} which are leading (zeroth) order in the recoil momentum $\vec{q}$ and DM-nucleus relative velocity $\vec{v}$. This is because WIMPs in the Galactic halo are highly non-relativistic ($v \sim 10^{-3} \mathrm{c}$), leading to typical momentum transfers in the 100 MeV range.

However, in recent years, a great deal of effort has been focused on developing a more general framework for DM-nucleus interactions. The non-relativistic effective field theory (NREFT; introduced by Fan et al. \cite{Fan:2010gt} and extended in Refs.~\cite{Fitzpatrick:2012ix,Anand:2013yka,Dent:2015zpa}) considers all possible non-relativistic quantum mechanical operators which can contribute to elastic DM-nucleus scattering, including those which are higher order in $\vec{q}$ and  $\vec{v}$. Such a basis of operators allows us to avoid biased reconstructions of future DM signals by accounting for all possible DM-nucleon interactions. In addition, it enables us explore possible blind spots in the sensitivity of current experiments. To this end, a number of authors have presented constraints on these operators from current and future direct detection \cite{Fitzpatrick:2012ib, Catena:2014uqa, Catena:2014epa, Schneck:2015eqa, Catena:2015uua, Scopel:2015baa} and neutrino telescope data \cite{Blumenthal:2014cwa, Catena:2015uha, Catena:2015iea, Vincent:2015gqa}. 

In this work, we consider the impact of NREFT operators on the recoil spectra in \textit{directional} direct detection experiments. As we discuss in detail in Sec.~\ref{sec:NREFT}, the event rates arising from NREFT operators typically appear with additional powers of $q^2$ and/or $v_\perp^2$ when compared with the standard SI/SD operators, where $\vec{v}_\perp$ is the DM velocity perpendicular to $\vec{q}$. Additional powers of $q^2$ typically result in a spectrum which grows with recoil energy $E_R$. When considering the directional rate integrated above a certain threshold energy $E_\mathrm{min}$, this enhances the contribution of directional features at high energy, leading to a more sharply directional event rate. In contrast, powers of $v_\perp^2$ tend to suppress recoils in the forward direction, leading to a more isotropic distribution. 

These differences can have significant consequences for interpreting future data from directional data. Calculations of the number of WIMP signal events required to discriminate from an isotropic background or to confirm the median recoil direction are performed assuming standard SI/SD interactions. As we will show, more general directional rates may require more or fewer events to confirm the WIMP origin of a signal. Additionally, some NREFT operators lead to distinctive directional signatures which may allow them to be distinguished from the standard scenario using directional detectors.

The key result of this paper is presented in Fig.~\ref{fig:Distinguish}, which shows the statistical significance with which standard interactions can be excluded as a function of the number of signal events. We show that certain NREFT operators (in particular $\mathcal{O}_7$ and $\mathcal{O}_8$, see Eq.~\ref{eq:operators}) can be distinguished from the standard SI/SD operators at the 95\% confidence level with several hundred signal events in a directionally-sensitive Fluorine-based detector. For these operators, discrimination would be almost impossible using only information about event energies, highlighting the important role directional detectors may play in exploring the particle physics of DM.

In Sec.~\ref{sec:Directional}, we describe in more detail the directional event rate. We then introduce the NREFT operators in Sec.~\ref{sec:NREFT} and demonstrate how their directional spectrum is expected to differ from the standard case. In Section~\ref{sec:StatisticalTests}, we consider a variety of directional statistics which have been proposed to distinguish a directional DM signal from isotropic backgrounds. Using these, we determine how many signal events are required to detect anisotropy and confirm the median recoil direction of the signal, and compare these results with the SI and SD rates which are typically considered. Finally, we discuss direct comparisons between the standard directional rate and these more general interactions. We show how directional detection can be used to distinguish non-relativistic operators which couple to the transverse velocity-squared from the canonical SI and SD operators.

\section{Directional rate}
\label{sec:Directional}
We begin with a summary of the standard event rate in directional detectors. The double-differential recoil spectrum per unit detector mass for DM-nucleus scattering with a fixed DM velocity $\vec{v}$ is given by \cite{Gondolo:2002np}:
\begin{equation}
\label{eq:Rate1}
\dbd{^2R}{E_R\mathrm {d}\Omega_q} = \frac{\rho_0 v}{2\pi m_\chi} \frac{\langle \left| \mathcal{M} \right|^2\rangle}{32 \pi m_N^2 m_\chi^2 v^2} \,\delta\left(\hat{v}\cdot \hat{q} - \vmin/v\right)\,.
\end{equation}
Here, $m_\chi$ and $m_N$ are the DM and nuclear masses respectively, $\rho_0$ is the local DM density and $\hat{q}$ is the direction of the recoiling nucleus. The matrix element-squared $\langle \left| \mathcal{M} \right|^2\rangle$, summed and averaged over the initial and final spins, is determined by the particle physics operators which mediate the interaction. Finally, the $\delta$-function imposes the kinematic constraint on the elastic scattering, where $\vmin$ is the minimum WIMP speed required to excite a nuclear recoil of energy $E_R$,
\begin{equation}
\vmin = \sqrt{\frac{m_N E_R}{2 \mu_{\chi N}^2}}\,,
\end{equation}
with the DM-nucleus reduced mass given by $\mu_{\chi N} = m_\chi m_N/(m_\chi + m_N)$. 

In the standard analysis framework, the matrix elements are calculated assuming contact interactions which are leading order in the momentum exchange and relative DM-nucleus velocity. This is because virialised DM particles in the Galactic halo are expected to have highly non-relativistic speeds, $v \sim 10^{-3} \mathrm{c}$, so any higher order interactions will be suppressed by factors of $10^{-6}$. These leading order interactions are assumed to arise from a coupling of the spins of the DM and nucleons (spin-dependent, SD) or a coupling of their particle densities (spin-independent, SI):
\begin{align}
\begin{split}
\label{eq:standardoperators}
\mathcal{O}_{SD} &= \vec{S}_\chi \cdot \vec{S}_n\\
\mathcal{O}_{SI} &= 1\,,
\end{split}
\end{align}
where we use the subscript $n$ for nucleon.

The resulting matrix elements are then written as
\begin{equation}
\langle \left| \mathcal{M} \right|^2\rangle =  \langle c^p \mathcal{O}^p + c^n \mathcal{O}^n\rangle\,,
\end{equation}
where $c^p$ and $c^n$ are the coupling strengths of the operators with protons and neutrons. The term in angular brackets  on the right hand side is referred to as the nuclear response function, and is the expectation value of the operators (either SI or SD) over all nucleons in the nucleus. For the SI case, this gives a coherent enhancement of the scattering rate roughly proportional to $A^2$, for a nucleus composed of $A$ nucleons. In the SD case, the response function takes into account the total nuclear spin, as well as the expectation values of the proton and neutron spin within the nucleus. For both types of interaction, the finite size of the nucleus leads to a loss of coherence at large momentum transfer, meaning that the nuclear response functions give a roughly exponential suppression of the rate with recoil energy \cite{Duda:2006uk, Cerdeno:2012ix}.

The final component for calculating the standard directional event rate arises from the fact that DM in the Galactic halo has a distribution of velocities $f(\vec{v})$. Thus, we must integrate Eq.~\ref{eq:Rate1} over all DM velocities, weighted by $f(\vec{v})$:
\begin{equation}
\label{eq:Rate2}
\dbd{^2R}{E_R\mathrm {d}\Omega_q} = \frac{\rho_0}{2\pi m_\chi} \frac{\langle \left| \mathcal{M} \right|^2\rangle}{32 \pi m_N^2 m_\chi^2} \, \hat{f}(\vmin, \hat{q})\,.
\end{equation}
All dependence on the velocity distribution has been absorbed into the Radon Transform (RT) \cite{Radon:17}, defined as:
\begin{equation}
\label{eq:RT}
\hat{f}(\vmin, \hat{q}) = \int_{\mathbb{R}^3} f(\vec{v}) \, \delta \left( \vec{v}\cdot \hat{q} - \vmin\right) \, \mathrm{d}^3\vec{v}\,,
\end{equation}
where we have changed variables in the argument of the $\delta$-function, leading to an extra power of $v$ in the integral. Physically, the RT is obtained by integrating over all velocities for which the observed recoil is kinematically allowed.

In the Standard Halo Model (SHM), dark matter is assumed to follow a Maxwell-Boltzmann velocity distribution, given by
\begin{equation}
\label{eq:SHM}
f(\vec{v}) = \frac{1}{(2\pi \sigma^2)^{3/2}} \exp\left[ -\frac{(\vec{v} - \vec{v}_\textrm{lag})^2}{2\sigma_v^2}  \right]\,.
\end{equation}
For an isotropic, isothermal sphere of DM, with density profile $\rho \propto r^{-2}$, the average velocity of the DM particles with respect to the Earth $\vec{v}_\textrm{lag}$ is related to the velocity dispersion by $v_\textrm{lag} = \sqrt{2}\sigma_v$. A value of $v_\textrm{lag} \approx 220 \kms$ is typically used \cite{Kerr:1986hz}, though values in the range $180-270 \kms$ have been suggested \cite{Feast:1997sb, Schonrich:2012qz, Bovy:2012ba, Lee:2013xxa, McCabe:2013kea}. The corresponding RT is given by \cite{Gondolo:2002np}:

\begin{equation}
\label{eq:SHMradon}
\hat{f}(\vmin, \hat{q}) = \frac{1}{(2\pi \sigma_v^2)^{1/2}} \exp\left[ - \frac{(\vmin - \vec{v}_\mathrm{lag}\cdot \hat{q})^2}{2\sigma_v^2} \right]\,.
\end{equation}

However, we note briefly that the SHM is unlikely to be a realistic description of the true DM distribution. Results from N-body simulations indicate deviations from the standard Maxwell-Boltzmann distribution \cite{Vogelsberger:2008qb, Butsky:2015pya}, including the possibility of additional structures such as dark disks \cite{Read:2009iv, Read:2009jy, Kuhlen:2013tra,Ruchti:2015bja} or streams \cite{Freese:2003na,Freese:2003tt}. For concreteness, we assume the SHM as a standard benchmark, with fixed values of $v_\mathrm{lag} = 220 \kms$ and $\sigma_v = 156 \kms$ in this study. We will also neglect effects due to the finite Galactic escape speed \cite{Piffl:2013mla}, which will not be significant over the range of recoil energies considered here. We leave an exploration of the impact of astrophysical uncertainties to future work. 

It will sometimes be necessary to distinguish the full double-differential recoil rate $\mathrm{d}^2R/{\mathrm{d}E_R \mathrm{d}\Omega_q}$ from the energy-integrated recoil rate, given by
\begin{equation}
\label{eq:AngRate}
\dbd{R}{\Omega_q} = \int_{E_\mathrm{min}}^{E_\mathrm{max}} \dbd{^2R}{E_R \mathrm{d}\Omega_q}\, \mathrm{d}E_R\,.
\end{equation}
In this case, we are interested in the direction of all recoils observed in the detectors (in an energy window $E \in [E_\mathrm{min}, E_\mathrm{max}]$), but not the energy of each event.

\section{Non-relativistic effective field theory (NREFT)}
\label{sec:NREFT}
In non-relativistic effective field theory (NREFT), the standard set of SI and SD operators are extended to include all those interactions constructed from Galilean, Hermitian and time-reversal invariant operators \cite{Fitzpatrick:2012ix}. This framework was extended in Ref.~\cite{Anand:2013yka} to include composite operators which do not typically arise due to the exchange of spin-0 or spin-1 mediators. The NREFT interaction operators are rotational-invariants constructed from the following Hermitian operators: 
\begin{equation}
i\frac{\vec{q}_n}{m_n}, \quad \vec{v}^{\perp}_n, \quad \vec{S}_\chi, \quad \vec{S}_n\,.
\end{equation}
Here, $\vec{q}_n$ is the momentum transferred to the interacting nucleon, $\vec{S}_{\chi, n}$ are the WIMP and nucleon spin operators and the operator $\vec{v}^{\perp}_n$ is defined as
\begin{equation}
\vec{v}^\perp_n = \vec{v} + \frac{\vec{q}_n}{2 \mu_{\chi n}}\,.
\end{equation}
This is the component of the DM velocity $\vec{v}$ perpendicular to the recoil momentum. By energy conservation, we therefore have $\vec{v}^\perp_n  \cdot \hat{q}_n = 0$. The DM velocity $\vec{v}$ does not have definite parity under the exchange of incoming and outgoing particles, and is therefore not Hermitian. The transverse velocity $\vec{v}^\perp_n$, however, is Hermitian and is therefore the only combination in which the DM velocity may appear. 

The full list of possible WIMP-nucleon operators is given in Appendix~\ref{sec:app:NREFT} (Eq.~\ref{eq:operators}), with notation matching that given in Refs.~\cite{Fitzpatrick:2012ix, Anand:2013yka}. Within this framework, we write the full interaction Lagrangian as
\begin{equation}
\label{eq:lagrangian}
\mathcal{L} = \sum_{i = 1}^{15} c_i^{0} \mathcal{O}_i^{0} + c_i^{1} \mathcal{O}_i^{1}\,,
\end{equation}
where the superscript indices $0,1$ denote the isoscalar and isovector couplings and operators respectively \footnote{In the proton-neutron basis, the couplings can be written $c^0 = \frac{1}{2}(c^p + c^n)$ and $c^1 = \frac{1}{2}(c^p - c^n)$.}. Within the NREFT framework, the standard SD and SI interactions are labeled $\Op{4}$ and $\Op{1}$ respectively, and have the exact form given in Eq.~\ref{eq:standardoperators}.

The matrix element-squared is then written as
\begin{equation}
\label{eq:matrixelements}
\langle \left| \mathcal{M} \right|^2\rangle = \sum_{i,j=1}^{15} \sum_{\tau, \tau' = 0,1}c_i^{\tau} c_j^{\tau'} F_{ij}^{\tau \tau'}(v_\perp^2, q^2)\,,
\end{equation}
where $F_{ij}$ are the nuclear response functions associated with the $i^\mathrm{th}$ and $j^\mathrm{th}$ operator.  For simplicity, we will neglect interference terms ($i \neq j$). In addition, we neglect the operator $\Op{2}$, which does not typically appear at leading order in the non-relativistic reduction of a relativistic interaction Lagrangian. The list of nuclear response functions in terms of a set of standard form factors is given in Eq.~\ref{eq:FormFactors} of Appendix~\ref{sec:app:NREFT}. We note that the standard SI and SD form factors are often normalised to unity at $q = 0$ \cite{Duda:2006uk}. However, in this framework, we include any coherent enhancement factors in the definition of the nuclear form factors.

We will also consider an example of a long-range DM-nucelon operator, as described in Refs.~\cite{Fornengo:2011sz,DelNobile:2013sia,Panci:2014gga}. Such operators are not contact operators but instead arise from the exchange of light mediator particles and are therefore not strictly effective field theory operators. However, we include an example in this work due to the novel $q^2$ dependence which they give rise to. The example we consider is $\mathcal{O}_{1}^{LR} = \mathcal{O}_1/q^2$ which behaves as $\mathcal{O}_1$ (the standard SI operator) with an additional $1/q^4$  suppression of the cross section.

We note that the response functions $F_{ij}$ depend on the incoming and outgoing momenta only through $q^2$ and $v_\perp^2$ \cite{DelNobile:2013sia}, which can be written as

\begin{equation}
v_\perp^2 = v^2 - \frac{q^2}{4\mu_{\chi N}^2}\,.
\end{equation}
These are now the WIMP-\textit{nucleus} momentum transfer and transverse velocity respectively. The operators forming the basis of the NREFT are rotationally invariant, so there is no preferred basis in which to measure the directions of $\vec{v}_\perp$ and $\vec{q}$. They can therefore only appear as scalar products. However, by construction $\vec{v}_\perp \cdot \vec{q} = 0$. Moreover, averaging over nuclear spins means that $\vec{S}_n$ does not pick out a particular direction. The result is that the response functions in Eq.~\ref{eq:matrixelements} depend only on the magnitude of the recoil momentum $q$, but not its direction $\hat{q}$. Substituting these response functions into Eq.~\ref{eq:Rate1} has no impact on the directionality of the signal; the only dependence on the recoil direction is through $\hat{v}\cdot\hat{q}$ within the $\delta$-function. 

However, in passing from Eq.~\ref{eq:Rate1} to Eq.~\ref{eq:Rate2}, care must be taken, because now the matrix elements can depend on the DM velocity, via their dependence on $v_\perp^2$. The full directional event rate, integrating over the WIMP velocity distribution, is then
\begin{align}
\begin{split}
\label{eq:RateEFT}
\frac{\mathrm{d}^2R}{\mathrm{d} E_R\mathrm {d}\Omega_q} &= \frac{\rho_0}{2\pi m_\chi} \frac{1}{32 \pi m_N^2 m_\chi^2} \sum_{i,j=1}^{15} \sum_{\tau, \tau' = 0,1}c_i^{\tau} c_j^{\tau'}\, \\
&\quad \times   \int_{\mathbb{R}^3} F_{ij}^{\tau \tau'}(v_\perp^2, q^2) f(\vec{v}) \, \delta \left( \vec{v}\cdot \hat{q} - \vmin\right) \, \mathrm{d}^3\vec{v}\,.
\end{split}
\end{align}
The response functions $F_{ij}$ are composed of terms which are proportional either to $v_\perp^0$ or $v_\perp^2$. In the former case, the integral over the velocity distribution is simply the Radon Transform of Eq.~\ref{eq:RT}. In the latter case, we must compute the transverse Radon Transform (TRT), which we define as
\begin{align}
\label{eq:modifiedRT}
\hat{f}^T(\vmin, \hat{q}) = \mathit{c}^{-2} \int_{\mathbb{R}^3} (\vec{v}_\perp)^2 f(\vec{v}) \,\delta(\vec{v}\cdot\hat{q} - \vmin) \, \mathrm{d}^3\vec{v}\,.
\end{align}
We have defined the TRT as carrying two inverse factors of the speed of light. In standard natural units, speeds are dimensionless, so this ordinarily does not need to be made explicit. However, we include this factor in order for the TRT to have the same units (of inverse speed) as the standard RT, to allow a more transparent comparison.

We note that for some operators (e.g.~$\mathcal{O}_7$), the relevant nuclear response function is proportional to $v_\perp^2$, meaning that the directional rate is proportional to the TRT of the velocity distribution. In other cases (e.g.~$\mathcal{O}_{10}$), the form factor has no dependence on $v_\perp$ and therefore the directional recoil rate (Eq.~\ref{eq:RateEFT}) has the same directional dependence as the standard SI/SD scenario. In general, however, both types of terms may be present and the full directional rate may be somewhere between the two regimes.

\subsection{Transverse Radon Transform}
 \label{sec:TRT}
 
In order to calculate the TRT, we can decompose the DM velocity into components perpendicular and parallel to $\hat{q}$,  $\vec{v} = (\vec{v}_\perp, v_{||})$, where we note that $\vec{v}_\perp$ is a two-dimensional vector. In this basis, we can write $\vec{v}\cdot \hat{q} = v_{||}$, meaning that Eq.~\ref{eq:modifiedRT} reduces to
\begin{align}
\label{eq:TRT}
\hat{f}^T(\vmin, \hat{q}) = \mathit{c}^{-2} \int_{\mathbb{R}^2} (\vec{v}_\perp)^2 f(\vec{v}_\perp, v_{||} = \vmin) \, \mathrm{d}^2\vec{v}_{\perp}\,.
\end{align}
The requirement that $v_{||} = \vmin$ ensures that the kinematic constraints of the elastic scattering are satisfied. It is then necessary to integrate over all possible transverse velocities, weighted by the transverse velocity squared. Geometrically, we must integrate $f(\vec{v})$ over a plane perpendicular to $\hat{q}$, at a distance $\vmin$ from the origin, weighted by the square of the perpendicular distance along the plane. For the SHM, given by Eq.~\ref{eq:SHM}, the TRT becomes

\begin{align}
\label{eq:SHM-int}
\hat{f}^T(\vmin, \hat{q}) &= \frac{1}{(2\pi \sigma^2)^{3/2}\mathit{c}^{-2}} \exp\left[ - \frac{(\vmin - \vec{v}_\mathrm{lag}\cdot \hat{q})^2}{2 \sigma_v^2} \right] \nonumber \\ 
&\quad\quad \times \int_{\mathbb{R}^2} (\vec{v}_\perp)^2 \, \exp\left[ - \frac{(\vec{v}_\perp - \vec{v}_\mathrm{lag}^\perp)^2}{2\sigma_v^2}\right] \mathrm{d}^2\vec{v}_{\perp}\,,
\end{align}
where $\vec{v}_\mathrm{lag}^\perp = \vec{v}_\mathrm{lag} - \vec{v}_\mathrm{lag}\cdot \hat{q}$. Performing the integral over transverse velocities, we obtain
\begin{align}
\label{eq:SHM-TRT}
\hat{f}^T(\vmin, \hat{q}) &= \frac{1}{(2\pi)^{1/2}\sigma_v \mathit{c}^2}\left(2\sigma_v^2 + v_\textrm{lag}^2 -  (\vec{v}_\textrm{lag}\cdot\hat{q})^2\right)  \nonumber \\ 
&\quad\quad \times \exp\left[ -\frac{(\vmin - \vec{v}_\textrm{lag}\cdot \hat{q})^2}{2\sigma_v^2}  \right]\,.
\end{align}

\begin{figure*}[tb!]
\centering
\includegraphics[width=0.45\textwidth]{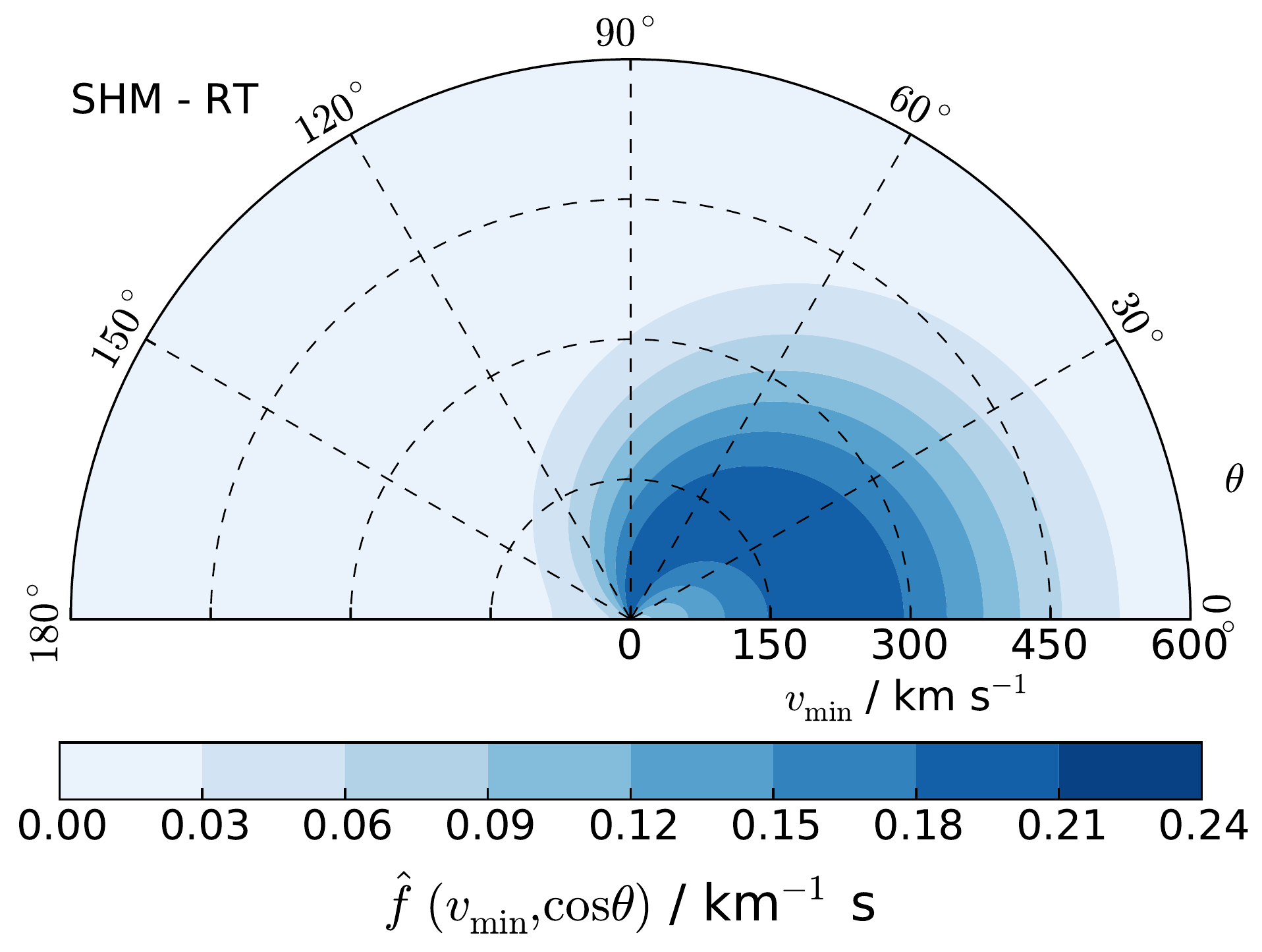}
\includegraphics[width=0.45\textwidth]{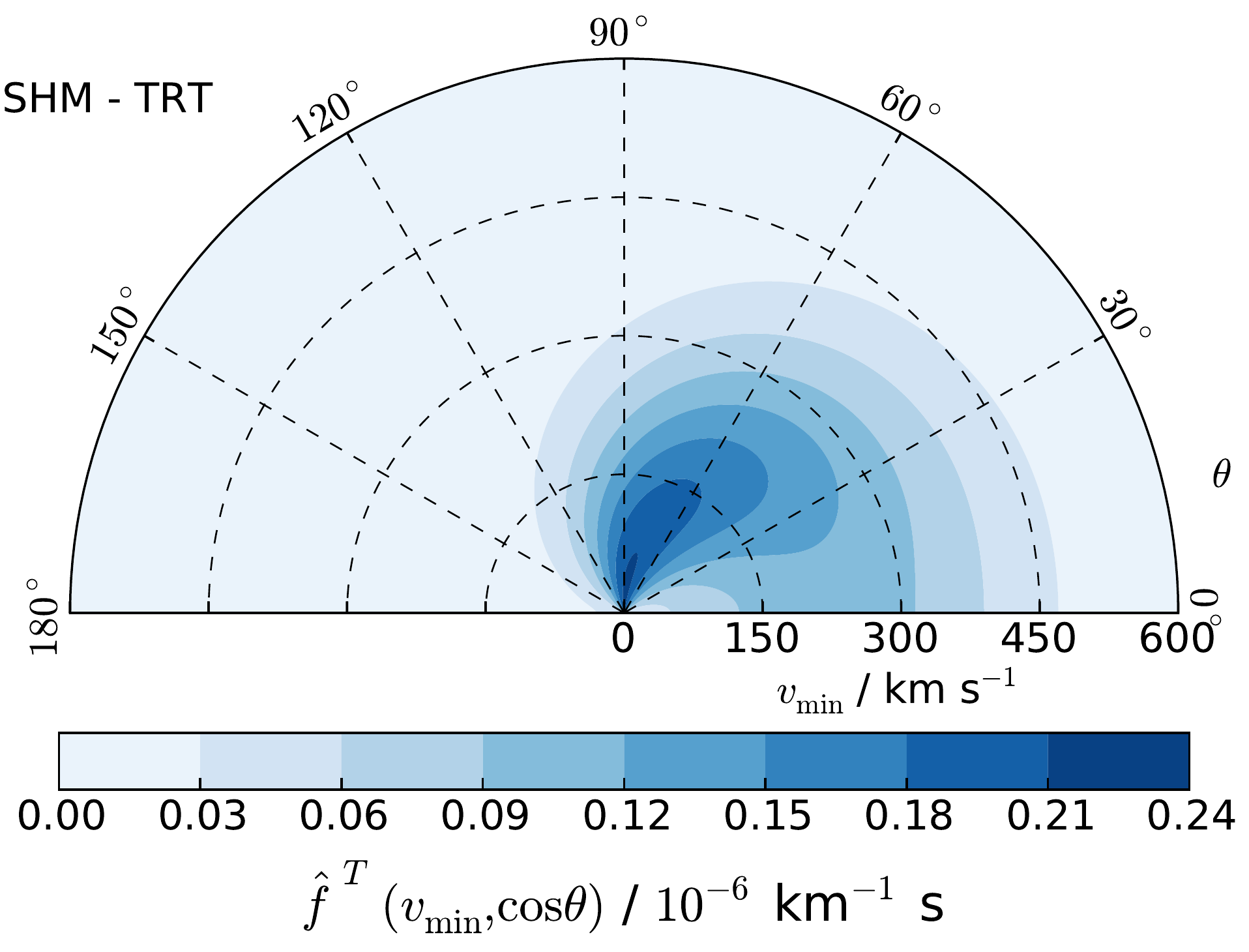}
\caption{\textbf{Comparison of standard and transverse Radon Transforms for the SHM.} The left panel shows the Radon Transform (RT) of the SHM (defined in Eq.~\ref{eq:SHM}), integrated over the azimuthal angle $\phi$ and with $\vec{v}_\mathrm{lag}$ aligned along $\theta = 0$. The right panel shows the corresponding transverse Radon Transform (TRT) which appears in the directional rate for NREFT operators coupling to $\vec{v}_\perp$ and is defined in Eq.~\ref{eq:modifiedRT}.}  
\label{fig:RT-comparison}
\end{figure*}

\begin{figure*}[tb!]
\centering
\includegraphics[width=0.45\textwidth]{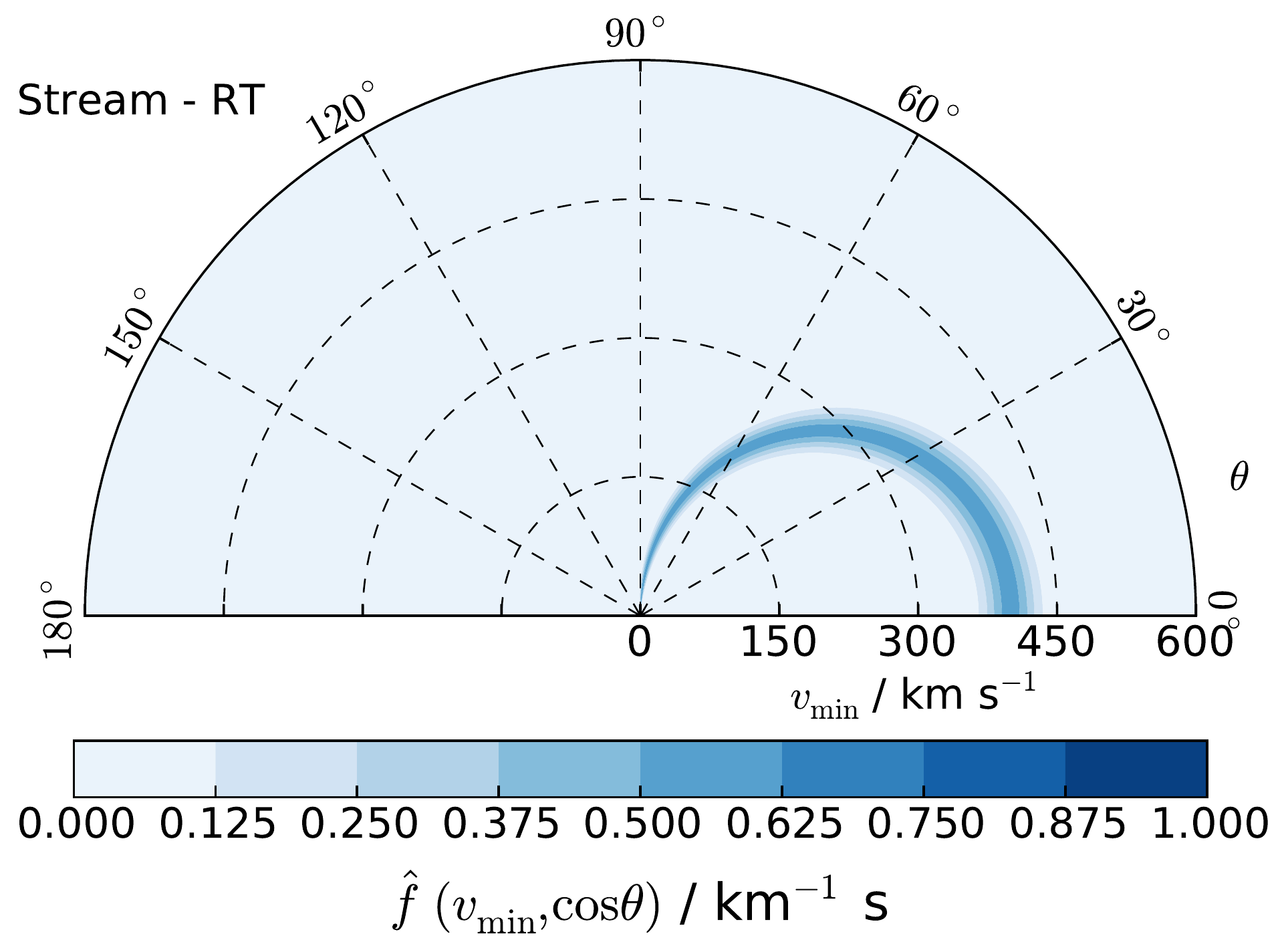}
\includegraphics[width=0.45\textwidth]{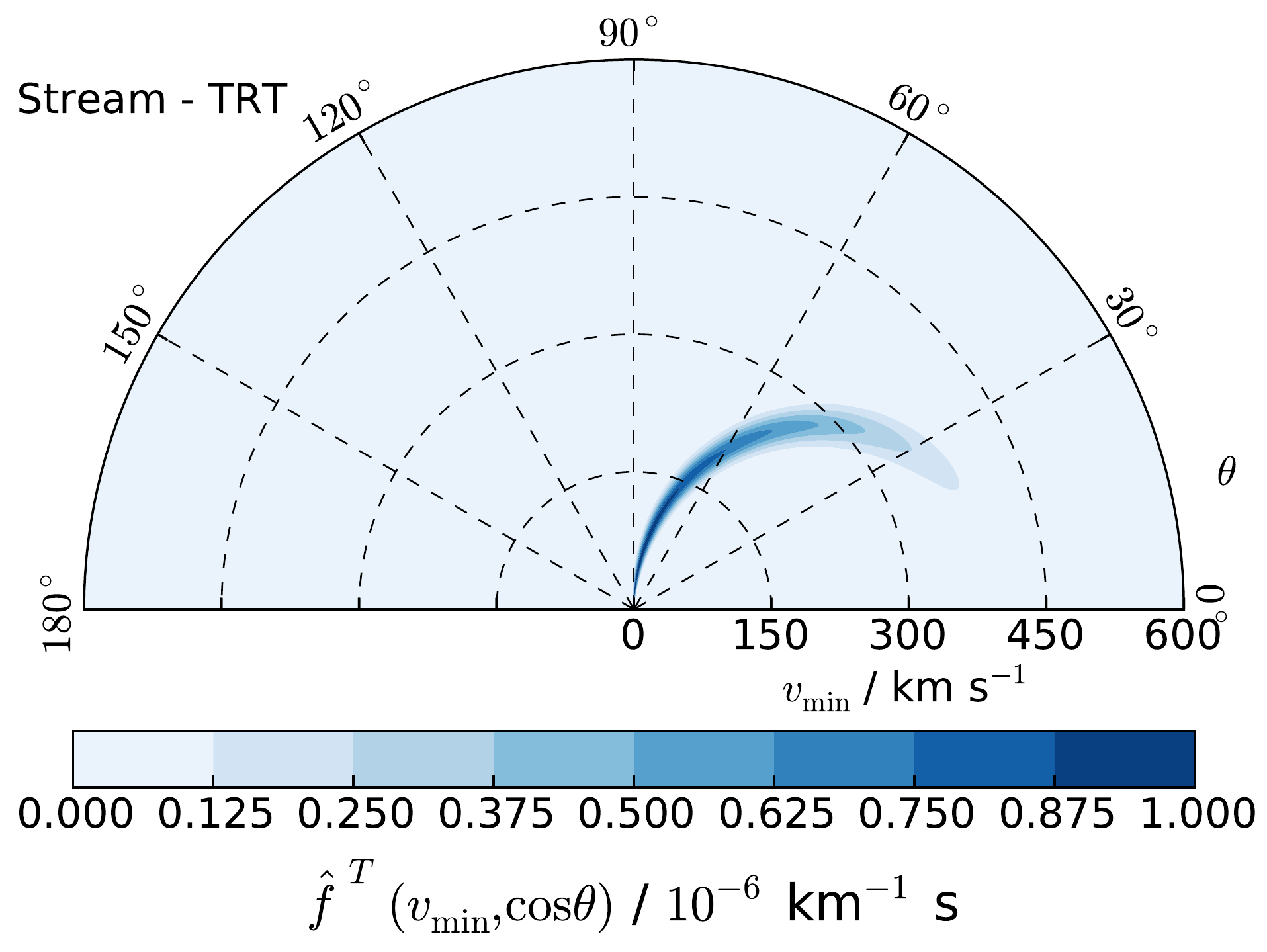}
\caption{\textbf{Comparison of standard and transverse Radon Transforms for a stream.} The left panel shows the Radon Transform (RT) of a stream distribution function, integrated over the azimuthal angle $\phi$ and with $\vec{v}_\mathrm{lag}$ aligned along $\theta = 0$. We approximate the stream as a Maxwell-Boltzmann distribution with $v_\mathrm{lag} = 400 \kms$ and $\sigma_v = 20 \kms$. The right panel shows the corresponding transverse Radon Transform (TRT) which appears in the directional rate for NREFT operators coupling to $\vec{v}_\perp$ and is defined in Eq.~\ref{eq:modifiedRT}.}\label{fig:RT-comparison-stream}
\end{figure*}

For more extreme distributions, such as a stream \cite{Freese:2003na, Freese:2003tt}, the velocity distribution can typically be modelled as a Maxwell Boltzmann distribution \cite{Peter:2011eu}  with a small velocity dispersion and $\vec{v}_\mathrm{lag} = \vec{v}_s$, the stream velocity. However, in the extreme case that $\sigma_v \rightarrow 0$, the stream distribution becomes
\begin{equation}
f(\vec{v}) = \delta^3\left(\vec{v} - \vec{v}_s\right)\,.
\end{equation}
The corresponding TRT in this case is then given by
\begin{equation}
\hat{f}^T(\vmin, \hat{q}) = \frac{\vec{v}_s^2 - (\vec{v}_s \cdot \hat{q})^2}{\mathit{c}^2}\,\delta\left(\vmin - \vec{v}_s\cdot\hat{q}\right)\,.
\end{equation}

In order to calculate the total energy spectrum of events, it is necessary to integrate the RT or TRT over all angles, depending on the relevant operator, in order to obtain the corresponding velocity integrals: 
\begin{align}
\label{eq:eta}
\eta(\vmin) &= \oint \hat{f} (\vmin,\hat{q}) \, \mathrm{d}\Omega_q\,, \\
\eta^T(\vmin) &= \oint \hat{f}^T (\vmin,\hat{q}) \, \mathrm{d}\Omega_q\,.
\end{align}
We have verified explicitly that the velocity integral  $\eta^T(\vmin)$ obtained in this way matches that given in Appendix A of Ref.~\cite{DelNobile:2013sia}.

In Fig.~\ref{fig:RT-comparison}, we compare the standard and transverse RTs for the SHM. For ease of presentation, we have integrated over the azimuthal direction $\phi$ and chosen the angular basis such that $\vec{v}_\mathrm{lag}$ is aligned along $\theta = 0$. The angle $\theta$ is therefore the angle between $\vec{v}_\mathrm{lag}$ and the recoil direction. The value of the TRT is approximately $10^{6}$ times smaller than the RT, as the TRT is suppressed by two powers of $v/\mathit{c} \sim 10^{-3}$. The most striking feature of the TRT, however, is that the maximum in the differential recoil rate does not occur along $\theta = 0$ as in the standard case. This is because forward-going WIMPs cannot induce forward-going nuclear recoils, due to the weighting by $v_\perp^2$. Instead, the maximum rate occurs approximately perpendicular to $\vec{v}_\mathrm{lag}$, where $v_\perp^2$ is maximised. However, the finite width of the SHM distribution means that a significant population of WIMPs will have velocities which deviate from $\vec{v}_\mathrm{lag}$. This means that recoils along the direction of $\vec{v}_\mathrm{lag}$ are still possible, so the forward scattering rate is not precisely zero.

For comparison, we show in Fig.~\ref{fig:RT-comparison-stream} the more extreme example of a stream, modelled as a Maxwell-Boltzmann distribution with $v_\mathrm{lag} = 400 \kms$ and $\sigma_v = 20 \kms$. As in the case of the SHM, the peak recoil direction deviates substantially from the forward direction. However, in the case of the stream, recoils in the forward direction are almost entirely suppressed. This is because of the very narrow velocity dispersion, which means that all particles are travelling with velocity close to $\vec{v}_\mathrm{lag}$ and so cannot induce recoils in that direction. The result is that, compared with the standard RT, the TRT is truncated at large values of $\vmin$, above around $350 \kms$. 

\subsection{Comparing NREFT operators}
\label{sec:OperatorComparison}

We now compare the directional rate obtained for the different NREFT operators. For concreteness, we will consider a CF$_4$ target, which is used in several gaseous TPC experiments \cite{Battat:2014van, Riffard:2015rga, Battat:2013gma, Miuchi:2011qw}, and which provides a promising SD WIMP-proton target. We will focus only on interactions with Fluorine. Carbon makes up only 14\% of CF$_4$ by mass and is spin-zero and we therefore expect the contribution from Carbon to be subdominant. We consider a WIMP of mass $m_\chi = 100 \,\, \mathrm{GeV}$ with SHM velocity distribution and an experimental sensitivity in the energy window $E_R \in [20, 50] \, \,\mathrm{keV}$. An energy threshold of 20 keV has previously been reported by the DRIFT-IId experiment \cite{Daw:2010ud}, although the angular resolution of directional experiments worsens at low energies \cite{Billard:2012bk}. We limit our analysis to isoscalar couplings ($c^p = c^n$), though as we will see, the differences in directionality arise predominant from the scaling of the different response functions with $v_\perp$ and $q$, so we do not expect the results to change substantially for more general couplings.

\begin{figure}[tb!]

\centering
\includegraphics[width=0.48\textwidth]{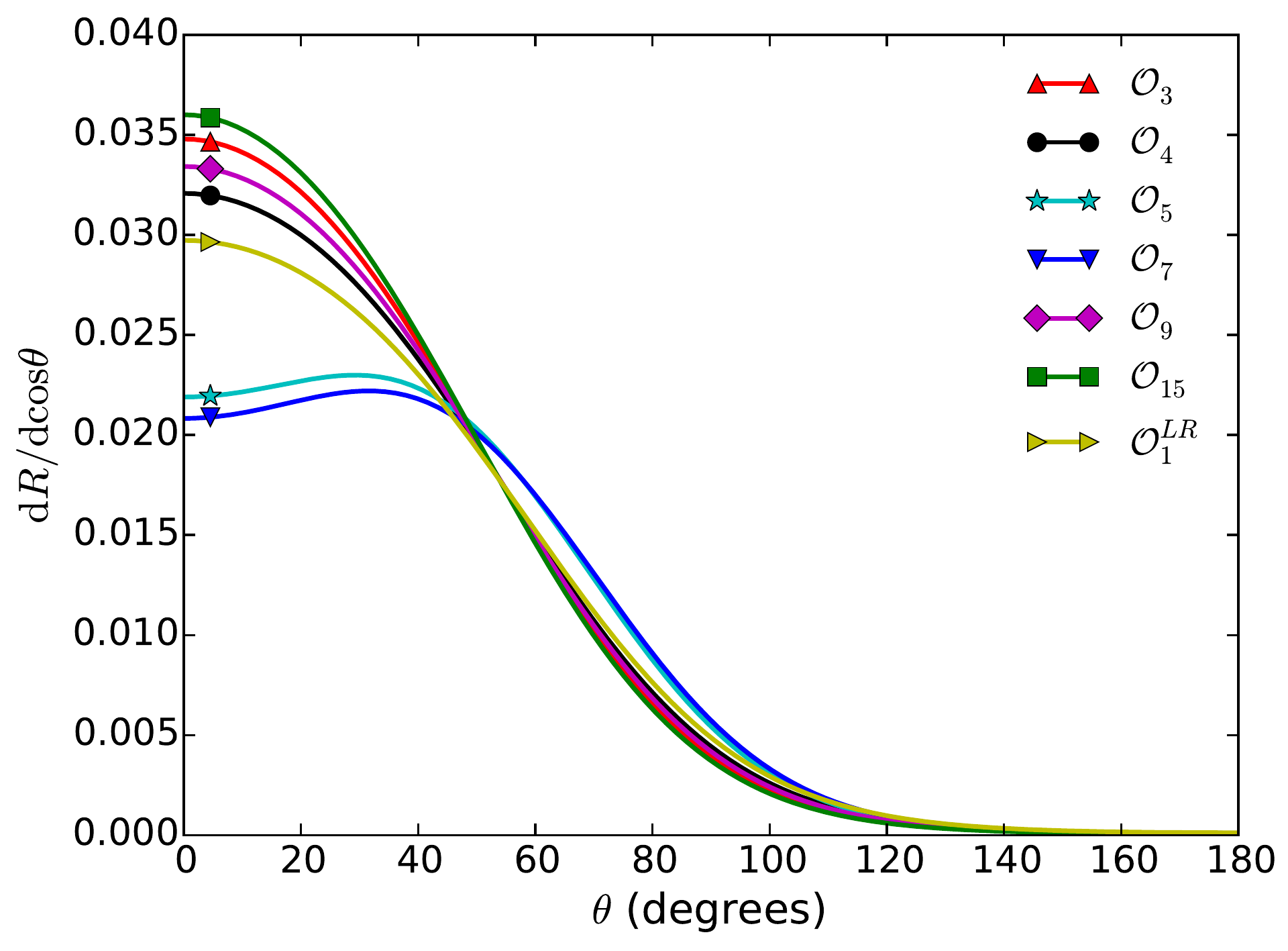}
\caption{\textbf{Directional event rates, normalised to unity, for several NREFT operators:} $\mathcal{O}_3$ (red $\blacktriangle$), $\mathcal{O}_4$ (black $\bullet$), $\mathcal{O}_5$ (cyan $\star$), $\mathcal{O}_7$ (blue $\blacktriangledown$), $\mathcal{O}_9$ (magenta $\blacklozenge$), $\mathcal{O}_{15}$ (green $\blacksquare$) and $\mathcal{O}_{1}^{LR}$ (yellow $\blacktriangleright$). The form of these operators is given in Eq.~\ref{eq:operators}. Each of the remaining NREFT operators gives a directional distribution similar to one of those shown here, depending on the functional dependence of the operator in question (see Eq.~\ref{eq:functional}). The angle $\theta$ is defined as the angle between $\vec{v}_\mathrm{lag}$ and the direction of the nuclear recoil. We assume $m_\chi = 100 \,\, \mathrm{GeV}$ and a Fluorine detector with rate integrated over $E_R \in [20, 50]$ keV.}
\label{fig:OperatorComparison}
\end{figure}

Figure~\ref{fig:OperatorComparison} shows the total directional rate integrated over the energy window of the experiment (defined in Eq.~\ref{eq:AngRate}) expressed as a function of $\theta$, the angle between $\vec{v}_\mathrm{lag}$ and the nuclear recoil direction. The directional rate has been normalised to unity to allow a comparison of the angular distribution of events between different NREFT operators. 

We show results only for a selection of operators. We find that each of the remaining operators leads to a directional rate which is almost indistinguishable from one of those plotted in Fig.~\ref{fig:OperatorComparison}. For light nuclei such as Fluorine, form factors do not decay as rapidly with $q$ as for heavier nuclei (and can often be assumed to be approximately constant \cite{Duda:2006uk}), so we expect that differences in form factors for the different operators should not be significant. This grouping of different operators therefore arises due to their different functional dependence on powers of $q^2$ and $v_\perp^2$. We classify the nuclear response functions for Fluorine as follows (in a similar fashion to the classification of Ref.~\cite{Anand:2013yka}): 
\begin{equation}
\label{eq:functional}
\textrm{Proportional to }
\begin{cases}
 1 &: \mathcal{O}_1, \mathcal{O}_4\,,\\
v_\perp^2 &: \mathcal{O}_7, \mathcal{O}_8\,,\\
q^2 &: \mathcal{O}_9, \mathcal{O}_{10}, \mathcal{O}_{11},\mathcal{O}_{12} \,, \\
v_\perp^2 q^2 &: \mathcal{O}_5, \mathcal{O}_{13}, \mathcal{O}_{14} \,,\\
q^4 &: \mathcal{O}_3, \mathcal{O}_{6} \,,\\
q^4 (q^2 + v_\perp^2) &:  \mathcal{O}_{15}\, ,\\
q^{-4} &: \mathcal{O}_1^{LR}\,.
\end{cases}
\end{equation}
Operators belonging to the same class will lead to approximately the same directional rate, so we therefore show only a single example from each class in Fig.~\ref{fig:OperatorComparison}.

The standard directional signal arising from SD (or SI) interactions corresponds to the operator $\mathcal{O}_4$ in Fig.~\ref{fig:OperatorComparison}. This standard signal lies intermediate between the remaining NREFT operators. Nuclear response functions suppressed by positive powers of $q^2$ lead to a directional rate more sharply peaked towards $\vec{v}_\mathrm{lag}$. This may be surprising, given that such response functions will affect only the energy dependence of Eq.~\ref{eq:RateEFT} and not the angular dependence (for a fixed value of $E_R$). However, the distribution of recoils becomes increasingly anisotropic with increasing $\vmin$, which can be seen from the left panel of Fig.~\ref{fig:RT-comparison}. When we integrate over all energies, response functions which scale as $q^2$ give a greater weight to this more anisotropic recoil distribution at high $\vmin$, leading to a more directionally peaked distribution. Equivalently, we note that by energy conversation, $q = 2 \mu_{\chi N} \vec{v}\cdot \hat{q} = 2 \mu_{\chi N} v \cos\theta$. This means that increasing the relative contribution of recoils at large $q$ is equivalent to increasing the contribution of recoils with small values of $\theta$, leading to a more peaked directional spectrum. Further powers of $q^2$ leads to increasingly forward-peaked directional spectra.

The long-range operator $\mathcal{O}_1^{LR}$ has a nuclear response function which scales as $q^{-4}$. Again, the directional dependence of the full double-differential rate is unchanged compared to the standard case. However, integrating over all energies, the greatest contribution now comes from low energy recoils, for which the spectrum is most isotropic. This leads to a directional spectrum which is less peaked in the forward direction (relative to the standard SI/SD interactions), as shown in Fig.~\ref{fig:OperatorComparison}. 

Nuclear response functions which are suppressed by powers of $v_\perp^2$ also lead to directional spectra which are less sharply peaked than the standard case. However, in this case, it is due to a fundamental difference in the directionality of the double-differential recoil spectrum. This behaviour is encoded in the TRT and arises because scattering in the forward direction is suppressed by the coupling to $v_\perp$, while scattering perpendicular to $\vec{v}_\mathrm{lag}$ is enhanced. In particular, for the operators $\mathcal{O}_7$ and $\mathcal{O}_5$ (and operators with similar functional forms) the peak in the recoil distribution occurs at some non-zero angle from the direction of peak flux $\vec{v}_\mathrm{lag}$. We can see this clearly in the right panel of Fig.~\ref{fig:RT-comparison}, where the TRT increases moving away from the forward recoil direction. For the operator $\mathcal{O}_5$, an additional suppression by $q^2$ means that the directional spectrum peaks at a slightly lower angle than in the case of $\mathcal{O}_7$. However, the result for both operators is that the recoil spectrum will exhibit a ring-like feature, with the peak recoil direction being greatest in a ring around the median recoil direction. 

\begin{figure}[t!]
\centering
\includegraphics[width=0.48\textwidth]{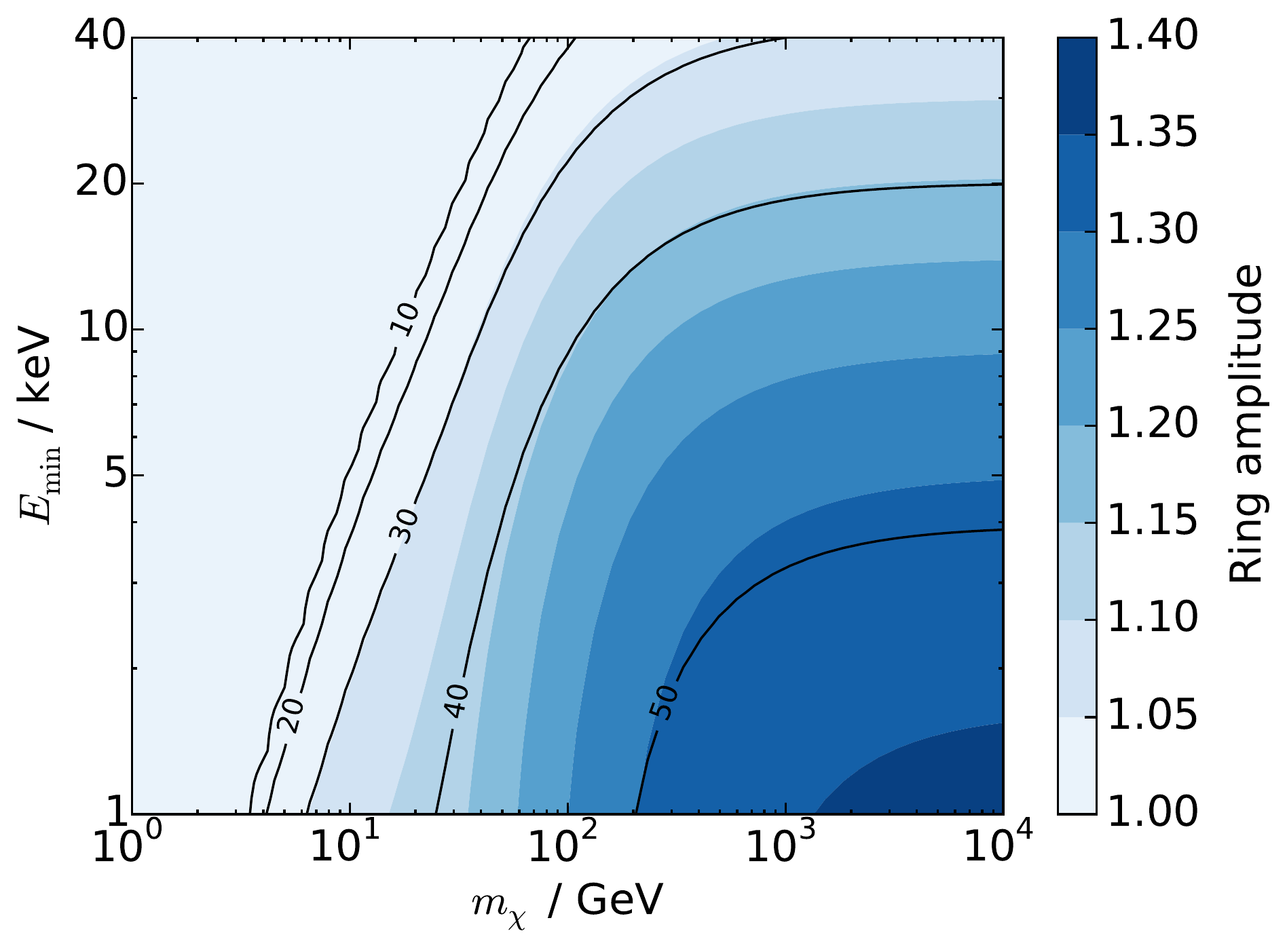}
\caption{\textbf{Properties of the ring in the directional recoil spectrum} (integrated over $E_R \in [E_\mathrm{min}, 50]$ keV) for the operator $\mathcal{O}_7$, assuming a Fluorine target. The ring opening angle in degrees is indicated by solid contours, while the ring amplitude (ratio of the maximum rate to the rate at $\theta = 0$) is shown by the blue shaded regions. The maximum ring opening angle for the parameters considered here is approximately $52^\circ$.}
\label{fig:ring}
\end{figure}

We illustrate the features of this ring in Fig.~\ref{fig:ring} for the operator $\mathcal{O}_7$, as a function of the WIMP mass and threshold energy. We continue to consider the directional spectrum integrated over recoil energies, defined in Eq.~\ref{eq:AngRate}. The solid contours indicate the ring opening angle in degrees (that is, the angle between $\vec{v}_\mathrm{lag}$ and the peak recoil direction). The shaded regions show the ring amplitude: the ratio between \dbd{R}{\theta} at the peak and at $\theta = 0$.

A similar ring-like feature in the standard directional rate has been studied previously \cite{Bozorgnia:2011vc}. When considering the standard RT, a ring is only possible if $\vmin < v_\mathrm{lag}$, meaning that the term inside the exponential in Eq.~\ref{eq:SHMradon} can be set to zero for some value of $\theta$. For the TRT, there is also a dependence on $\theta$ outside the exponential. By differentiating Eq.~\ref{eq:SHM-TRT} with respect to $\theta$, it can be shown that a maximum in $\hat{f}^T$ for $\theta \neq 0$ therefore exists for larger values of $\vmin$ than in the standard case, up to $\vmin = 2 v_\mathrm{lag}$. This means that the ring arising from NREFT operators which couple to $v_\perp^2$ can be observed for lower WIMP masses, higher energy thresholds and for smaller values of $v_\mathrm{lag}$. For example, for the range of parameter values displayed in Fig.~\ref{fig:ring}, no significant ring-like feature is observable in the directional spectrum (integrated over energy) for the standard SI or SD operators, giving a maximum ring opening angle of $\sim 1^\circ$ (compared to $52^\circ$ for the $\mathcal{O}_7$ operator).\footnote{A significant ring is found in Ref.~\cite{Bozorgnia:2011vc} for larger values of $v_\mathrm{lag}$ than those assumed here or, alternatively, when the full spectrum is considered (i.e. when we do not integrate over energy).} A detailed study of ring-like features is conducted in Ref.~\cite{Bozorgnia:2011vc}, but we note here that if a prominent ring is observed for low mass WIMPs or for a relatively high threshold, this may be indicative of non-standard operators, which couple to the transverse WIMP speed.

In general then, the basis of NREFT operators leads to a variety of directional signatures. Those operators suppressed by powers of $q^2$ lead to a more anisotropic directional spectrum (when we consider event directions, integrating over all event energies), while those which couple to $v_\perp^2$ give a more isotropic spectrum, with the possibility of a ring-like feature in the spectrum, even for relatively large values of $\vmin$. 

In the rest of this work, we focus on the operators $\mathcal{O}_7$ and $\mathcal{O}_{15}$, which are written,

\begin{align}
\mathcal{O}_7 &= \vec{S}_n \cdot \vec{v}_\perp \,,\\
\mathcal{O}_{15} &= -(\vec{S}_\chi \cdot \frac{\vec{q}}{m_n})((\vec{S}_n \times \vec{v}_\perp)\cdot \frac{\vec{q}}{m_n})\,,
\end{align}
and lead to the following nuclear response functions,
\begin{align}
\label{eq:NuclearResponse}
F_{7,7} &\propto v_\perp^2  F_{\Sigma'} \,,\\
F_{15,15} &\propto \frac{q^4}{m_n^4} \left( v_\perp^2 F_{\Sigma'}   + 2 \frac{q^2}{m_n^2}F_{\Phi''} \right)\,.
\end{align}
As can be seen in Fig.~\ref{fig:OperatorComparison}, the operator $\mathcal{O}_7$ leads to the most isotropic directional rate, while the operator $\mathcal{O}_{15}$ leads to the most sharply peaked directional rate. We note in the case of $\mathcal{O}_{15}$ that there are two terms in the nuclear response function, one proportional to $q^6$ and the other proportional to $q^4 v_\perp^2$. These give similar contributions, as the overall normalisations of $F_{\Sigma'}$ and $F_{\Phi''}$ are similar, meaning that the recoil spectrum of $\mathcal{O}_{15}$ is expected to differ strongly from the standard case in both energy and directional spectra. These two example distributions represent the most extreme departures from the standard SI/SD operator case and allow us to explore the full range of behaviours of the NREFT operators.

\section{Statistical tests}
\label{sec:StatisticalTests}
In this section, we explore how the two operators $\Op{7}$ and $\Op{15}$ differ from the standard directional rate with regards to two statistical tests which have been proposed to confirm the DM nature of a directional signal. First, we explore how many events are required to reject the isotropy of the signal. Second, we determine how many events are required to confirm the median recoil direction of the signal. This allows us to quantify the particle physics uncertainties associated with the signals, arising from a lack of knowledge about which NREFT operators mediate the WIMP-nucleon interaction.

We use the same experimental parameters as in Sec.~\ref{sec:OperatorComparison} and assume perfect energy resolution and angular resolution. Of course, realistic experiments have finite energy resolution and are expected to have an angular resolution in the range $20^\circ - 80^\circ$, depending on the recoil energy \cite{Billard:2012bk}. However, this idealised case allows us to place a lower limit on the number of events required to distinguish the DM signal from an isotropic background. Furthermore, our focus is on comparing different operators, and assuming an idealised experiment allows us to disentangle experimental uncertainties from effects arising from varying particle physics.

\subsection{Rejecting isotropy}
\label{sec:isotropy}

Backgrounds of terrestrial origin are expected to be isotropically distributed, so the first step in confirming the WIMP origin of a signal is to reject isotropy of the signal. We follow Morgan \textit{et al}.~\cite{Morgan:2004ys} and use the modified Rayleigh-Watson statistic $\mathcal{W}^\star$, defined in Ref.~\cite{Mardia:2002}, as a measure of isotropy. Large values of $\mathcal{W}^\star$ indicate a larger degree of anisotropy. We calculate $\mathcal{W}^*$ from the directions of mock events (distributed according to Eq.~\ref{eq:Rate2}), discarding information about event energies. In order to determine the number of signal events $N_\mathrm{WIMP}$ required to reject isotropy, we use the following procedure for each value of $N_\mathrm{WIMP}$:
\begin{enumerate}
\item Generate 10000 mock data sets, each consisting of $N_\mathrm{WIMP}$ recoil directions, distributed assuming a particular NREFT operator;
\item Calculate $\mathcal{W}^\star_\mathrm{WIMP}$ for each mock data set;
\item Calculate the 5\% percentile of $\mathcal{W}^\star_\mathrm{WIMP}$;
\item Generate a further 10000 mock data sets, each consisting of $N_\mathrm{WIMP}$ recoil directions, distributed isotropically;
\item Calculate $\mathcal{W}^\star_\mathrm{iso}$ for each mock data set;
\item Calculate the 95\% percentile of $\mathcal{W}^\star_\mathrm{iso}$.
\end{enumerate}
The 95\% percentile of $\mathcal{W}^\star_\mathrm{iso}$ is the value of $\mathcal{W}^\star$ above which we would reject isotropy at the 95\% confidence level. We then find the value of $N_\mathrm{WIMP}$ for which this value is equal to the 5\% percentile of $\mathcal{W}^\star_\mathrm{WIMP}$.  For this value of $N_\mathrm{WIMP}$, we can expect to reject isotropy at the 95\% level in 95\% of experiments in which the signal events are distributed according to the NREFT operator of interest. 

\begin{figure}[tb!]
\centering
\includegraphics[width=0.48\textwidth]{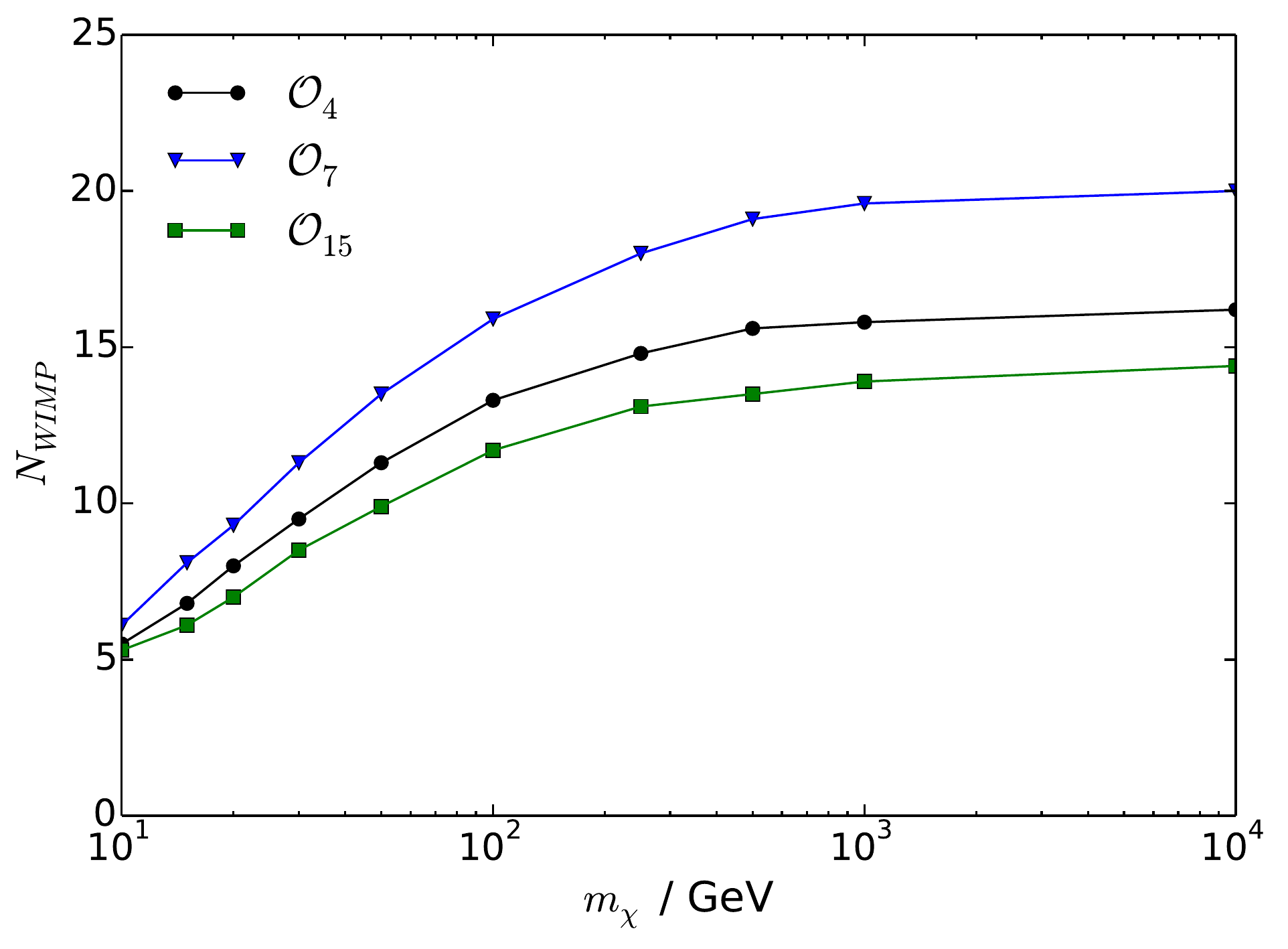}
\caption{\textbf{Number of WIMP signal events required to reject isotropy at 95\% confidence.} Results are shown assuming the signal is distribution according to each of three different NREFT operators: $\mathcal{O}_4$ (black $\bullet$), $\mathcal{O}_7$ (blue $\blacktriangledown$) and $\mathcal{O}_{15}$ (green $\blacksquare$). A Fluorine-based detector with an energy threshold of 20 keV is assumed.}
\label{fig:Isotropy}
\end{figure}

Figure~\ref{fig:Isotropy} shows the results for the three operators $\Op{4}$ (the standard SD operator), $\Op{7}$ and $\Op{15}$ as a function of the WIMP mass. The number of events required to reject isotropy, assuming the standard SD operator, is of order 10, in good agreement with the previous results of Morgan \textit{et al}.~\cite{Morgan:2004ys}. As expected, $\Op{15}$ requires fewer events to reject isotropy than the standard operators, while $\Op{7}$ requires more. For example, at $m_\chi = 100 \,\, \mathrm{GeV}$, the standard operator $\Op{4}$ requires 12 events to reject isotropy, compared to 16 and 11 events for $\Op{7}$ and $\Op{15}$ respectively. Though this difference is relatively small in absolute terms, it represents an uncertainty of around 25\% in the number of events required, arising entirely from particle physics uncertainties.

At low masses, the number of events required for the three operators converges, with all three requiring only $\sim 6$ events for a WIMP mass of 10 GeV. This is because at low WIMP masses, only large values of $\vmin$ contribute to the event rate. This means that the exponential terms in Eqs.~\ref{eq:SHMradon} and \ref{eq:SHM-TRT} will decay rapidly away from the forward direction, leading to highly directional rates for all three operators. However, as we increase the WIMP mass, the number of events required for each operator begins to diverge. In particular, $N_\mathrm{WIMP}$ rises more rapidly for the operator $\Op{7}$. As demonstrated in Fig.~\ref{fig:ring}, the maximum in the directional spectrum moves further from $\theta = 0$ as we increase $m_\chi$, leading to a more isotropic distribution of events and therefore more signal events required to reject isotropy.

\subsection{Confirming median recoil direction}
\label{sec:median}
Once the isotropy of the signal has been confirmed, it will then be necessary to determine whether the median recoil direction matches that expected from a WIMP signal. For all three operators we consider, the expected median recoil direction is the same and is in the direction of $\vec{v}_\mathrm{lag}$. However, the distribution of observed median recoil directions over an ensemble of experiments will be different. In order to quantify this, we follow Ref.~\cite{Green:2010zm} and examine the distribution of $\Delta$, defined as the angle between the observed median recoil direction and the direction of Solar motion.

\begin{figure}[tb!]
\centering
\includegraphics[width=0.48\textwidth]{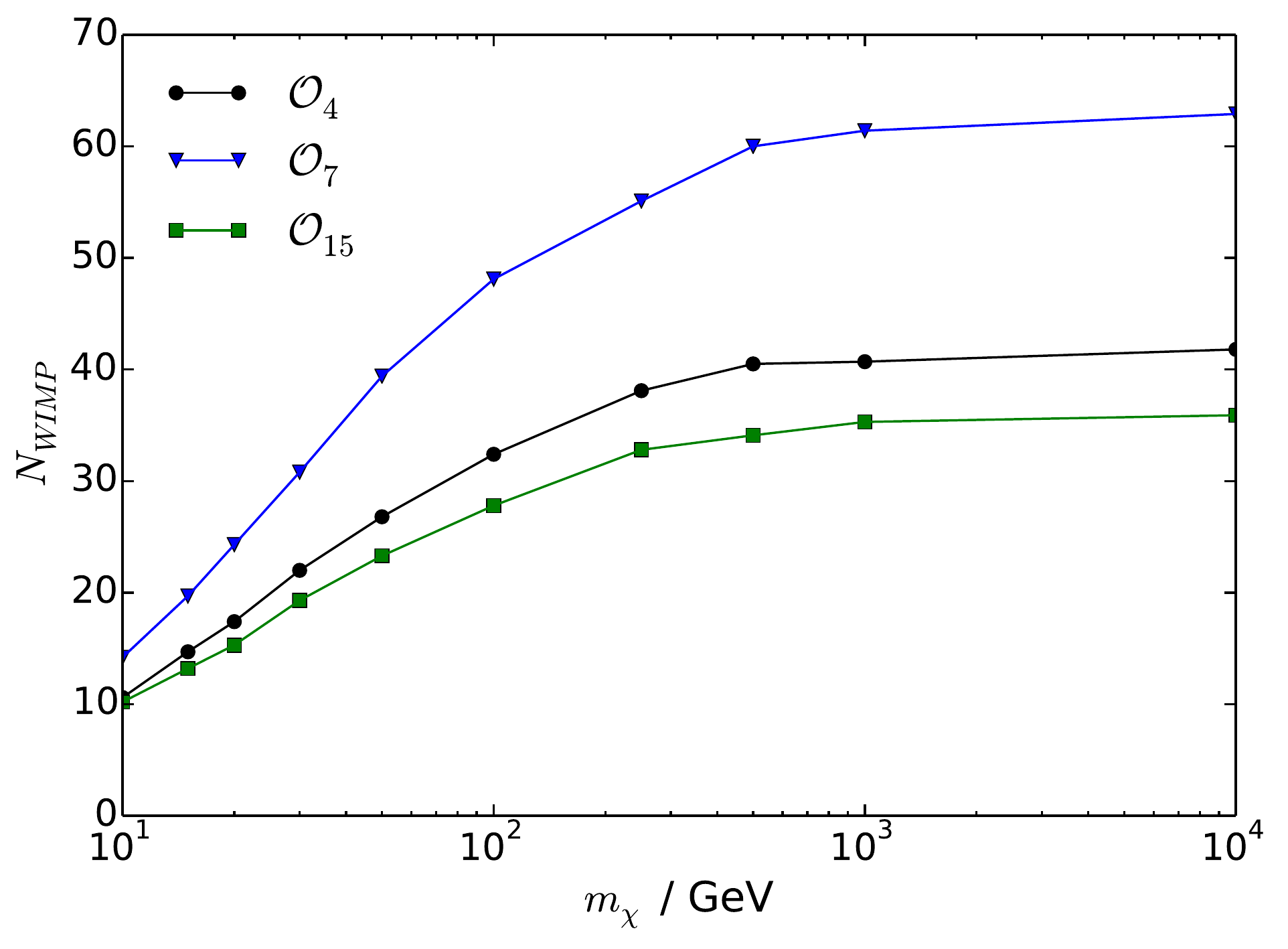}
\caption{\textbf{Number of WIMP signal events required to confirm median recoil direction at 95\% confidence.} Results are shown assuming the signal is distribution according to each of three different NREFT operators: $\mathcal{O}_4$ (black $\bullet$), $\mathcal{O}_7$ (blue $\blacktriangledown$) and $\mathcal{O}_{15}$ (green $\blacksquare$). A Fluorine-based detector with an energy threshold of 20 keV is assumed.}
\label{fig:Median}
\end{figure}

As in Sec.~\ref{sec:isotropy}, we generate 10000 mock data sets for each hypothesised signal, as well as for the null hypothesis of isotropic recoils, and calculate the distribution of $\Delta$. We then calculate the value of $N_\mathrm{WIMP}$ for which the 5\% percentile of $\Delta$ under the null hypothesis matches the 95\% percentile of $\Delta$ under the signal hypothesis. The results are shown in Fig.~\ref{fig:Median}.

As in the case of rejecting isotropy, the results agree with those presented previously \cite{Green:2010zm}, with around 30 events required to confirm the median recoil direction. Once again, $\Op{4}$ and $\Op{15}$ give almost identical results at low masses, as both have a highly directional recoil spectrum. In this case, however, $\Op{7}$ requires a larger number of events, approximately 50\% more at low masses, increasing substantially as the WIMP mass is increased. Even at low mass, where the peak recoil direction coincides with $\theta = 0$, the directional spectrum is broadened by the structure of the transverse Radon Transform, arising from the coupling to $v_\perp^2$. This increases the size of fluctuations in the median recoil direction away from $\theta = 0$. As the WIMP mass is increased, the ring-like structure described in Sec.~\ref{sec:OperatorComparison} becomes significant and more events are required to confirm the median recoil direction. Above around 1000 GeV, the number of events required is around a factor of 2 higher than for the other two operators.

While our results for standard interactions are in agreement with previous results, we have demonstrated that for NREFT operators, differences in directional spectra lead to different numbers of signal events required to confirm the median recoil direction. For a 100 GeV WIMP, this required number of events ranges between 25 and 50, effectively introducing a factor of 2 particle physics uncertainty into otherwise model independent statements about the WIMP origin of a signal.

\section{Comparison with standard interactions}
\label{sec:Distinguish}

Finally, we consider the possibility of discriminating between NREFT operators and standard SI/SD operators. In Sec.~\ref{sec:StatisticalTests}, we considered relatively model-independent statistical tests which allowed us to distinguish between signal and background. This was possible because the background hypothesis had a fixed form, that of isotropically distributed recoils. When comparing two signal hypotheses, however, this form is no longer fixed. This is because the signal rate for a given operator depends on the WIMP mass, which we assume is \textit{a priori} unknown. The signal rate may also depend on other uncertainties, such as in the astrophysical distribution of WIMPs, or in detector performance. However, we neglect these uncertainties in the present study, focusing on the idealised, fixed-astrophysics case. 

In order to make statistically robust statements then, we must compare the observed distribution of events with that expected from each operator \textit{for all possible values of the WIMP mass}. In order to account for this, and to make use of as much information as possible, we perform a full likelihood-ratio analysis \cite{Cowan:2010js}. The null hypothesis $H_0$ asserts that the signal is due entirely to standard SD interactions. The alternative hypothesis $H_1$ is that there is some contribution from another one of the NREFT operators (either $\Op{7}$ or $\Op{15}$, which we consider one at a time). The fraction of signal events due to one of these NREFT operators is denoted $A$, while the remaining fraction $(1-A)$ of events arises from the standard SD operator. The likelihood $L(m_\chi, A)$ is then the probability of obtaining the observed event energies and directions for a given value of $m_\chi$ and $A$.

For each of the 10000 mock data sets, we generate $N_\mathrm{WIMP}$ events, distributed in energy and direction according to either $\Op{7}$ or $\Op{15}$. We then calculate the following test statistic:
\begin{equation}
\label{eq:LikeRatio}
q_0 =  -2\ln \left[ \frac{L(\hat{\hat{m}}_\chi, 0)} {L(\hat{m}_\chi, \hat{A})}  \right]\,.
\end{equation}
Here, $L(\hat{\hat{m}}_\chi, 0)$ is the likelihood under the null hypothesis (i.e. SD-only events), maximised over all values of $m_\chi$. The unconditional maximum likelihood is then denoted $L(\hat{m}_\chi, \hat{A})$, where we maximise over both the WIMP mass $m_\chi$ and the non-standard operator fraction $A$. According to Wilks' theorem \cite{Wilks:1938dza}, for data distributed under the null hypothesis (SD interactions-only), $q_0$ is asymptotically $\chi^2$-distributed with one degree of freedom (the difference in dimensionality between the null and alternative hypotheses).\footnote{We have verified this numerically for several values of $N_\mathrm{WIMP}$ above 30.} 

For each value of $N_\mathrm{WIMP}$, we calculate $q_0^{95\%}$, defined such that 95\% of experiments observe a value of $q_0$ greater than or equal to $q_0^{95\%}$. We can then calculate the $p$-value (and corresponding confidence level) for the SD-only hypothesis, based on this value, as
\begin{equation}
\label{eq:pvalue}
p = \int_{q_{0}^{95\%}}^\infty P(\chi^2_1) \, \mathrm{d}\chi^2_1\,,
\end{equation}
where $P(\chi^2_1)$ is the probability density function for the $\chi^2_1$ distribution. That is, we calculate the probability of observing a value of $q_0$ as large or larger than $q_0^{95\%}$ if all signal events are due to standard SD interactions. A $p$-value this small or smaller will then be obtained in 95\% of experiments in which the signal is due to non-standard interactions. The smaller the value of $p$, the greater the confidence level with which we can reject the SD-only hypothesis and infer the presence of other NREFT operators.

\begin{figure}[tb!]
\centering
\includegraphics[width=0.48\textwidth]{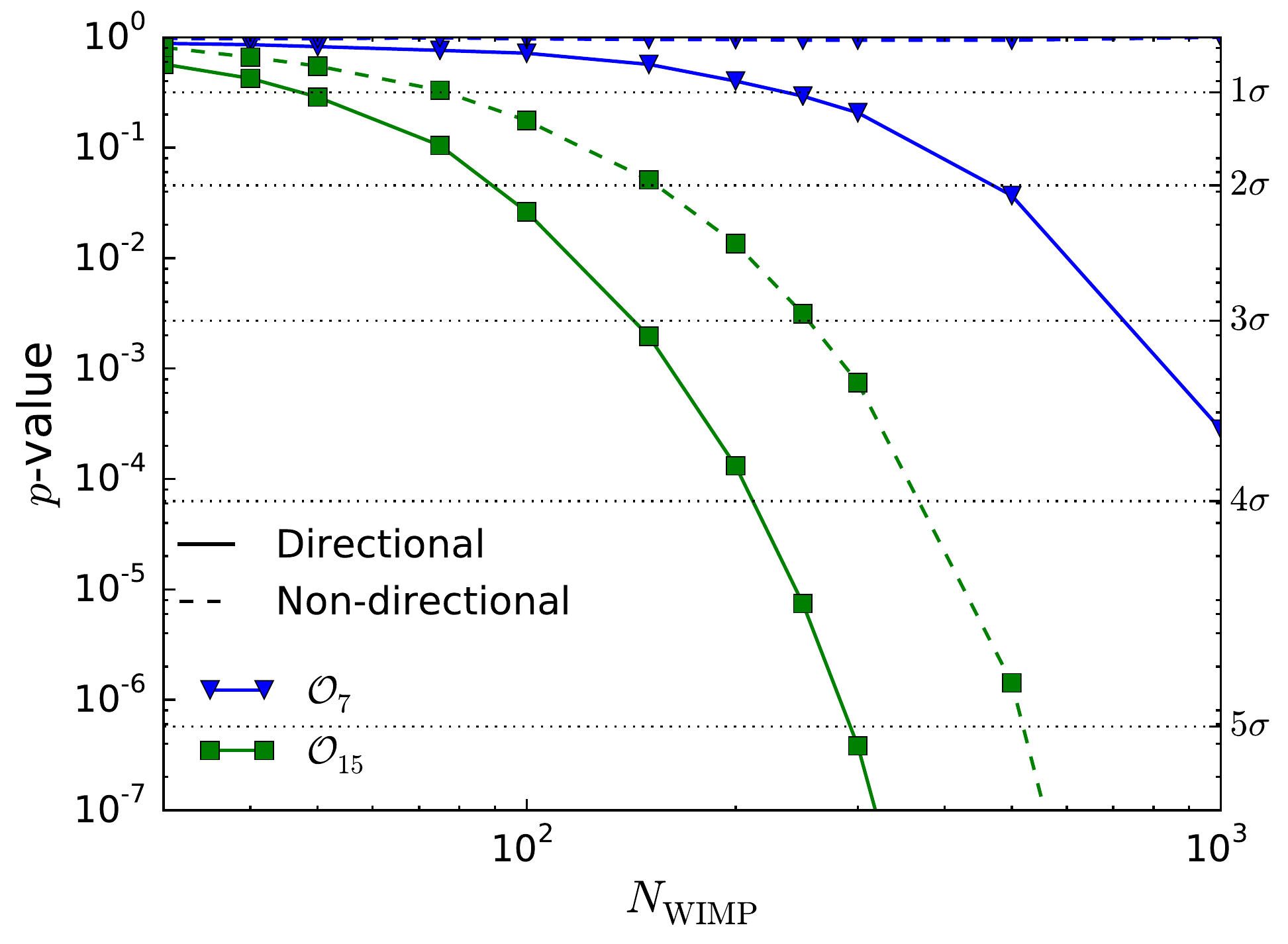}
\caption{\textbf{Confidence levels for rejecting standard SD-only interactions.} We show the $p$-value obtained in 95\% of experiments as a function of the number of signal events $N_\mathrm{WIMP}$ for the null hypothesis of standard SD-only interactions. We also show the corresponding confidence level with which standard interactions can be rejected (in units of $\sigma$). The signal is distributed according to the operators $\mathcal{O}_{7}$ (blue $\blacktriangledown$) or $\mathcal{O}_{15}$ (green $\blacksquare$), with a WIMP mass of 100 GeV. A Fluorine-based detector with an energy threshold of 20 keV is assumed. Experiments with and without directional information are shown as solid and dashed lines respectively. Note that the curve for $\mathcal{O}_{7}$ using only non-directional information lies above $p > 0.9$ for all values of $N_\mathrm{WIMP}$ considered.}
\label{fig:Distinguish}
\end{figure}

The resulting $p$-values and confidence levels (in units of $\sigma$) are shown in Fig.~\ref{fig:Distinguish} as a function of $N_\mathrm{WIMP}$, assuming signals arising from $\mathcal{O}_7$ (blue) and $\mathcal{O}_{15}$ (green), with a WIMP mass of 100 GeV. We show the results obtained when the full energy and directional information is used to calculate the likelihood (solid lines) as well as those obtained using energy information only (dashed lines). 

\begin{figure}[tb!]
\centering
\includegraphics[width=0.48\textwidth]{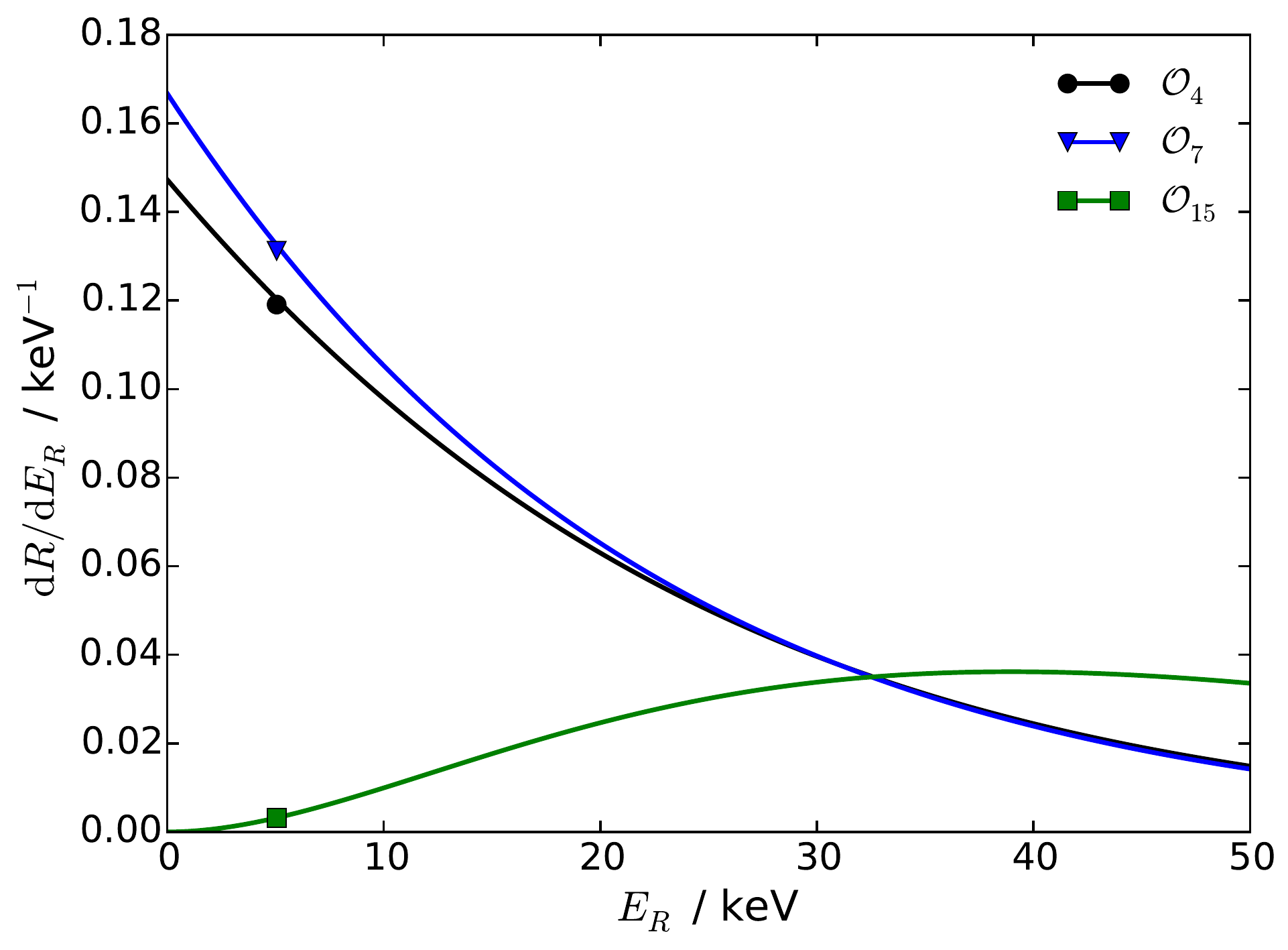}
\caption{\textbf{Energy spectra, normalised to unity, for several NREFT operators considered in Sec.~\ref{sec:Distinguish}:} $\mathcal{O}_4$ (black $\bullet$), $\mathcal{O}_7$ (blue $\blacktriangledown$) and $\mathcal{O}_{15}$ (green $\blacksquare$). The energy spectra are normalised to a single event in the range $E_R \in [20, 50]$ keV. We assume $m_\chi = 100 \,\, \mathrm{GeV}$ and a Fluorine detector.}
\label{fig:Spectra}
\end{figure}

In the case of non-directional detection, the $p$-value obtained for $\mathcal{O}_7$ remains large ($p > 0.9$), even with up to 1000 signal events. Figure~\ref{fig:Spectra} shows the energy spectra for the operators considered in this section and illustrates that the spectra for the standard SD ($\mathcal{O}_4$) and $\mathcal{O}_7$ operators are almost identical. As described in Sec.~\ref{sec:OperatorComparison}, for light nuclei such as Fluorine, we expect differences between the different form factors to be negligible. The most significant difference between the two spectra is therefore that the velocity integral for  $\mathcal{O}_7$ is $\eta^T(\vmin)$, defined in Eq.~\ref{eq:eta}, which is weighted by $v_\perp^2$. 

The behaviour of $\eta^T(\vmin)$ can be understood by examining Eq.~\ref{eq:SHM-TRT}. If $\sigma_v$ is large compared to $v_\mathrm{lag}$, then the exponential term in the TRT will vary slowly as a function of $\vec{v}_\mathrm{lag} \cdot \qhat = v_\mathrm{lag} \cos\theta$. When we integrate over all angles, the term proportional to $\vec{v}_\mathrm{lag} \cdot \qhat$ is then sub-dominant compared to the remaining terms, meaning that the velocity integral $\eta^T(\vmin)$ will be roughly proportional to the standard integral $\eta(\vmin)$. If instead $\sigma_v$ is small, the exponential will be sharply peaked and the angular integral will be dominated by $\vec{v}_\mathrm{lag}\cdot \qhat = \vmin$. In this case, the term proportional to $\vec{v}_\mathrm{lag} \cdot \qhat$ can be comparable to the remaining terms and will give an extra contribution to the energy spectrum proportional to $-\vmin^2$. 

Physically, a small value of $\sigma_v$ leads to a narrow velocity distribution. This means that almost all WIMPs are travelling with velocity close to $\vec{v}_\mathrm{lag}$. Due to the coupling to $\vec{v}_\perp^2$, scattering in the forward direction is suppressed, meaning that scattering through larger angles (and therefore with lower recoil energies) is enhanced. This is seen in the right panel of Fig.~\ref{fig:RT-comparison-stream} and was previously discussed in Sec.~\ref{sec:TRT}. For larger values of $\sigma_v$, there is a significant population of WIMPs travelling with large velocities at an angle to $\vec{v}_\mathrm{lag}$. These can induce high energy recoils in the direction of $\vec{v}_\mathrm{lag}$ while still satisfying the kinematic constraint $\vec{v}\cdot \qhat = \vmin$, meaning that the energy spectrum at high $E_R$ is not depleted. This can be seen in particular in the forward direction in the right panel of Fig.~\ref{fig:RT-comparison}. In the case of the SHM, with $v_\mathrm{lag} = \sqrt{2}\sigma_v$, this effect means that the energy spectrum obtained from the TRT matches the standard case within a few percent. Such small differences can easily be compensated for by varying the WIMP mass, making discrimination difficult with energy-only detectors.

When directional information is included, however, the distribution of events can be distinguished. As is clear from Fig.~\ref{fig:OperatorComparison}, the transverse Radon Transform leads to a different angular distribution of events for $\Op{7}$ when compared with the standard RT for the SD interaction. These distributions are sufficiently different that with around 500 events, the standard interactions can be rejected at the $2\sigma$ level in 95\% of experiments which are directionally sensitive, with $3\sigma$ discrimination possible with around 700 events. We emphasise that in this case, energy-only information does not allow us to significantly distinguish between the two operators. Thus, it is only the directionality of the signal which allows us to discriminate and reject the SD-only hypothesis, in favour of the NREFT operator $\Op{7}$.

By contrast, energy-only experiments can very quickly distinguish a signal dominated by $\Op{15}$ from a standard SD-only signal. The $2\sigma$ level is reached in 95\% of experiments with around 150 signal events, while the $5\sigma$ discovery level could be achieved with as few as 550 events. This is because of the characteristic energy spectrum produced by $\Op{15}$, which rises as $q^6$ for small $q$, before form factor suppression becomes important and the spectrum flattens at high energy. This spectrum is also illustrated in Fig.~\ref{fig:Spectra} and cannot be easily mimicked by the standard SD signal, even if the WIMP mass is varied.

When directional information is also included, again we see a significant improvement in the confidence level with which standard interactions can be rejected. For a given number of signal events, the rejection of standard interactions is approximately $1\sigma$ more significant when directional information is included. As a result, a $5\sigma$ rejection of SD-only scattering can be achieved for around 300 signal events. This arises because the nuclear response function for $\mathcal{O}_{15}$, shown in Eq.~\ref{eq:NuclearResponse}, contains a contribution proportional to $q^4 v_\perp^2$ as well as a contribution which goes as $q^6$. This means that as well as producing a different energy spectrum, $\mathcal{O}_{15}$ also produces a different directional distribution. Because the two contributions to the nuclear response function have a similar normalisation, this difference in directionality can be easily observed and can be used to distinguish $\mathcal{O}_{15}$ from the standard SD-only case.

\section{Discussion}

In this work, we have considered the directional recoil spectra produced by the set of NREFT operators. We have focused on a single, idealised Fluorine-target detector for concreteness. Though we have assumed a reasonable energy threshold of 20 keV \cite{Daw:2010ud}, we have assumed perfect angular resolution (compared to the typical resolution of $20^\circ$-$80^\circ$ \cite{Billard:2012bk}). We have also assumed that there is no background contamination. Introducing a finite angular resolution and background would increase the number of events required to reject isotropy and confirm the median recoil direction. The numbers reported above in Sec.~\ref{sec:StatisticalTests} therefore represent a lower limit and illustrate that even with an idealised detector, uncertainties coming from particle physics can be as much as a factor of 2. 

As well as experimental uncertainties, these results are subject to astrophysical uncertainties. Though we have presented the TRT for a stream distribution for illustration purposes in Fig.~\ref{fig:RT-comparison-stream}, we have restricted calculations to the SHM with fixed parameters. In Ref.~\cite{Morgan:2004ys}, the impact of different halo models on the number of events required to reject isotropy was studied. The values obtained therein vary by around 20\%, meaning that the particle physics uncertainties presented here are expected to be comparable to astrophysical uncertainties.

We note in particular that one of the conclusions of Sec.\ref{sec:Distinguish} - that $\Op{7}$ and the standard SD operator $\Op{4}$ are indistinguishable without directional information - is only true for SHM-like velocity distributions. In cases where $\sigma_v$ is significantly smaller than $v_\mathrm{lag}$ (such as in the presence of a stream), the recoil spectra for the two operators will diverge. However, this is also likely to accentuate the differences between the angular event distributions for the two operators, meaning that we expect that directional sensitivity will still provide a significant improvement in discrimination. The impact of astrophysical uncertainties in the NREFT framework has been briefly discussed in the past \cite{Scopel:2015baa}, though clearly a detailed study of such uncertainties in directional and non-directional experiments will be necessary in future.

Even if the velocity integrals for the operators $\Op{4}$ and $\Op{7}$ are indistinguishable, discrimination between these different operators may also be possible through other methods. For heavier nuclei, such as Xenon, there may be more significant differences in the form factors associated with each operator. By comparing the energy spectra and number of events in several experiments using different target nuclei, it may be possible to determine which form factor (and therefore which operator) is mediating the interaction (see e.g.~\cite{Schneck:2015eqa}). However, many NREFT operators lead to interactions with the same form factor. In addition, uncertainties in calculations of the form factors and in the value of the WIMP mass may make such an approach more difficult \cite{Cerdeno:2012ix,Cerdeno:2013gqa}. A more promising approach is to measure the annual modulation of the dark matter signal, which has a different time-dependence for operators coupling to $\vec{v}_\perp$. However, the annual modulation would have to measured in several experiments and compared before different operators could be discriminated \cite{DelNobile:2015tza}. The use of directional information instead allows operators to be distinguished with just a single experiment.  

Finally, we caution that the likelihood-based approach of Sec.~\ref{sec:Distinguish} would not be appropriate for all NREFT operators. As previously discussed in Sec.~\ref{sec:OperatorComparison}, those operators which differ from the standard SI/SD interactions only through $q^2$ suppression do not change the directional dependence of the full double-differential recoil distribution (Eq.~\ref{eq:RateEFT}), but only affect the energy spectrum of events. Thus, the addition of directional information does not improve prospects for discrimination compared to the energy-only case, as the directional dependence of both operators is the same. In spite of this, the statistical tests of Sec.~\ref{sec:StatisticalTests} are still useful for such operators, as they allow robust and relatively model-independent comparisons to be made with isotropic backgrounds, without any reference to the energy spectrum of the operator.

\section{Conclusions}

In the current work, we have explored the directional signatures of DM-nucleon interactions within the framework of non-relativistic effective field theory (NREFT). Some of the operators arising in NREFT lead to a suppression by the recoil momentum $q^2$ and therefore can affect the angular distribution of recoils when the directions (but not energies) of the events are considered. Other operators lead to a coupling to the WIMP velocity perpendicular to the recoil direction $v_\perp^2$. In these cases, the directionality of recoils can be fundamentally different compared to the standard SI and SD interactions. A number of the NREFT operators give similar directional recoil spectra and can be classified, as in Eq.~\ref{eq:functional}, according to the scaling of their cross sections with $q^2$ and $v_\perp^2$. 

We have focused on two operators in particular, namely
\begin{align}
\mathcal{O}_7 &= \vec{S}_n \cdot \vec{v}^\perp \,, \textrm{ and} \\
\mathcal{O}_{15} &= -(\vec{S}_\chi \cdot \frac{\vec{q}}{m_n})((\vec{S}_n \times \vec{v}^\perp)\cdot \frac{\vec{q}}{m_n})\,.
\end{align}
The operator $\Op{15}$ produces recoils which are more strongly peaked in the forward direction, compared to the SI/SD case. The operator $\Op{7}$ instead produces recoils which are the most isotropic of all the NREFT operators. In Eq.~\ref{eq:TRT}, we have defined the transverse Radon Transform (TRT), which takes into account the $v_\perp^2$-weighting of the interaction cross section, and which appears in the recoil spectrum for $\Op{7}$ as well as several other operators. The TRT suppresses scattering in the forward direction and can therefore produce a pronounced ring-like feature in the directional recoil spectrum. Though ring-like features have previously been discussed in the context of standard operators \cite{Bozorgnia:2011vc}, the ring arising from the TRT should be observable down to lower WIMP masses ($m_\chi \lesssim 20 \,\, \mathrm{GeV}$) and for higher threshold energies (as high as 20-30 keV for high mass WIMPs). 

In Sec.~\ref{sec:StatisticalTests}, we have explored the number of events required for each operator to reject the isotropy of the recoils and to confirm the median recoil direction. For the strongly directional operator $\Op{15}$, only a slightly smaller number of recoils is required compared with standard SD interactions. For $\Op{7}$, however, substantially more recoils are required, due to the broader recoil distribution in this case, with the difference increasing as we consider higher WIMP masses. For a 100 GeV WIMP, we conclude that 10-15 events are required to reject isotropy, while 25-50 events are required to confirm the median recoil direction, depending on the operator in question. 

In Sec.~\ref{sec:Distinguish}, we considered how well the NREFT operators could be distinguished from the standard case using both directional and non-directional information. For an underlying $\Op{15}$ signal, SD-only interactions could be rejected at the $3\sigma$ level with around 300 events if only energy information is available. Including information about the recoil directions, this number is reduced to around 150 events. For $\Op{7}$, the energy spectrum of events cannot be distinguished from the standard case. However, when differences in the angular distribution of events are taken into account, it may be possible to exclude the SD-only scenario at the $3\sigma$ level with around 700 events. Though such large numbers of events would require large exposures (and are certainly beyond the scope of current experiments), future experiments with larger target masses and lower energy thresholds may be able to distinguish the different operators. 

We have demonstrated that directional information may be the only means of discriminating those operators which couple to $\vec{v}_\perp$ from those which do not. Though this study is far from exhaustive (neglecting, for example, interference terms between different operators), we have highlighted the importance of directional detection for probing the particle physics nature of Dark Matter.

During the preparation of this manuscript, a pre-print also discussing the directional rates in NREFT was made available online (Ref.~\cite{Catena:2015vpa}). In that paper, the author considers the directional spectra and relative contributions of different NREFT operators for several possible target materials in directional detectors. Instead, we have considered a single target material (CF$_4$) and focused on comparing the directional spectra produced by each operator. We have also considered the possibility of distinguishing between different operators using directional detection. However, the results of this paper and Ref.~\cite{Catena:2015vpa} are in broad agreement, including our expressions for the Radon Transforms and our predictions of a novel ring-like signature for certain NREFT operators.

\section*{Acknowledgments}
BJK would like to thank Paolo Panci for spirited and helpful discussions on both NREFT operators and on directional detection, as well as thanking Anne Green for helpful comments on this manuscript. The author also acknowledges the hospitality of the Institut d'Astrophysique de Paris, where part of this work was done. BJK is supported by the European Research Council ({\sc Erc}) under the EU Seventh Framework Programme (FP7/2007-2013)/{\sc Erc} Starting Grant (agreement n.\ 278234 --- `{\sc NewDark}' project). 

\appendix
\section{Non-relativistic operators}
\label{sec:app:NREFT}

Here, we list the NREFT operators which are considered in this work. At the nucleon level, they are constructed from the following Hermitian operators: the momentum transfer $ i\vec{q}/m_n$; the transverse WIMP-nucleon velocity $\vec{v}^{\perp}$; the DM spin $\vec{S}_\chi$; and the nucleon spin $\vec{S}_n$. The transverse velocity is given by 

\begin{equation}
\vec{v}_\perp = \vec{v} + \frac{\vec{q}}{\mu_{\chi n}}\,,
\end{equation}
where $\mu_{\chi n} = m_\chi m_n/(m_\chi + m_n)$ is the WIMP-nucleon reduced mass and $m_n$ is the nucleon mass. The list of NREFT operators is then as follows \cite{Fitzpatrick:2012ix, Anand:2013yka, DelNobile:2013sia}:
\begin{align}
\begin{split}
\label{eq:operators}
\mathcal{O}_1 &= 1 \\
\mathcal{O}_3 &= i \vec{S}_n \cdot (\frac{\vec{q}}{m_n} \times \vec{v}^{\perp})\\
\mathcal{O}_4 &= \vec{S}_\chi\cdot\vec{S}_n\\
\mathcal{O}_5 &= i  \vec{S}_\chi \cdot (\frac{\vec{q}}{m_n} \times \vec{v}^{\perp})\\
\mathcal{O}_6 &= (\vec{S}_\chi \cdot \vec{q})(\vec{S}_n \cdot \vec{q})\\
\mathcal{O}_7 &= \vec{S}_n \cdot \vec{v}^\perp\\
\mathcal{O}_8 &= \vec{S}_\chi \cdot \vec{v}^\perp\\
\mathcal{O}_9 &= i \vec{S}_\chi \cdot (\vec{S}_n \times \vec{q}) \\
\mathcal{O}_{10} &= i \vec{S}_n \cdot \vec{q} \\
\mathcal{O}_{11} &= i \vec{S}_\chi \cdot \vec{q} \\
\mathcal{O}_{12} &= \vec{S}_\chi \cdot (\vec{S}_n \times \vec{v}^\perp)\\
\mathcal{O}_{13} &= i(\vec{S}_\chi \cdot \vec{v}^\perp)(\vec{S}_n \cdot \frac{\vec{q}}{m_n})\\
\mathcal{O}_{14} &= i(\vec{S}_\chi \cdot \frac{\vec{q}}{m_n})(\vec{S}_n\cdot\vec{v}^\perp)\\
\mathcal{O}_{15} &= -(\vec{S}_\chi \cdot \frac{\vec{q}}{m_n})((\vec{S}_n \times \vec{v}^\perp)\cdot \frac{\vec{q}}{m_n})\,.
\end{split}
\end{align}
We neglect the operator $\Op{2} = \vec{v}_\perp^2$, as it does not arise at leading order from a relativistic Lagrangian without significant cancellation. It is therefore typically sub-dominant to other operators in the list. We also omit the two operators recently reported in Ref.~\cite{Dent:2015zpa}. Several dictionaries are available which allow one to translate from a relativistic interaction Lagrangian to the NREFT operators listed above \cite{Fitzpatrick:2012ix,DelNobile:2013sia,Dent:2015zpa}.

In addition, we have considered an example of a long-range operator:
\begin{equation}
\mathcal{O}_1^{LR} = \frac{\mathcal{O}_1}{q^2}\,.
\end{equation}
This operator behaves as $\mathcal{O}_1$, with an additional $q^{-4}$ suppression of the nuclear response function.

From the nucleon-level operators, it is necessary to calculate the matrix elements of these operators within the nucleus, summing over the contributions of all nucleons. Neglecting interference terms between different operators, the resulting nuclear response functions are given by:
\begin{align}
\begin{split}
\label{eq:FormFactors}
F_{1,1} &= F_{M}\,, \\
F_{3,3} &= \frac{1}{8}\frac{q^2}{m_n^2} \left( v_\perp^2 F_{\Sigma'} + 2\frac{q^2}{m_n^2} F_{\Phi''} \right)\,, \\
F_{4,4} &= \frac{C(j_\chi)}{16}\left( F_{\Sigma'} + F_{\Sigma''}\right)\,, \\
F_{5,5} &= \frac{C(j_\chi) }{4}  \frac{q^2}{m_n^2}  \left( v_\perp^2 F_{M} +  \frac{q^2}{m_n^2} F_{\Delta} \right)\,, \\
F_{6,6} &= \frac{C(j_\chi)}{16} \frac{q^4}{m_n^4} F_{\Sigma''}\,,\\
F_{7,7} &= \frac{1}{8} v_\perp^2  F_{\Sigma'}\,, \\
F_{8,8} &=  \frac{C(j_\chi)}{4} \left(  v_\perp^2 F_{M} +  \frac{q^2}{m_n^2} F_{\Delta} \right) \,,\\
F_{9,9} &=  \frac{C(j_\chi)}{16} \frac{q^2}{m_n^2} F_{\Sigma'}\,,\\
F_{10,10} &= \frac{1}{4}\frac{q^2}{m_n^2} F_{\Sigma''}\,,\\
F_{11,11} &= \frac{1}{4}\frac{q^2}{m_n^2} F_M\,,\\
F_{12,12} &= \frac{C(j_\chi)}{16} \left( v_\perp^2 \left( F_{\Sigma''} + \frac{1}{2}F_{\Sigma'}\right) + \frac{q^2}{m_n^2} \left(F_{\tilde{\Phi}'} + F_{\Phi''}\right)  \right)\,,\\
F_{13,13} &= \frac{C(j_\chi)}{16} \frac{q^2}{m_n^2} \left(  v_\perp^2 F_{\Sigma''}  + \frac{q^2}{m_n^2}F_{\tilde{\Phi}'}  \right)\,,\\
F_{14,14} &= \frac{C(j_\chi)}{32} \frac{q^2}{m_n^2} v_\perp^2 F_{\Sigma'} \,,\\
F_{15,15} &= \frac{C(j_\chi)}{32} \frac{q^4}{m_n^4} \left( v_\perp^2 F_{\Sigma'}   + 2 \frac{q^2}{m_n^2}F_{\Phi''} \right)\,.
\end{split}
\end{align}
Here, $C(j_\chi) = 4 j_\chi (j_\chi + 1)/3$, where $j_\chi$ is the DM spin. The transverse velocity appearing here is the (complex-valued) WIMP-nucleus velocity:
\begin{equation}
\vec{v}_\perp = \vec{v} + \frac{\vec{q}}{\mu_{\chi N}}\,,
\end{equation}
with $N$ denoting the \textit{nuclear}, rather than \textit{nucleon}, mass. We have suppressed the isospin indices ($\tau, \tau'$) which appear in Eq.~\ref{eq:matrixelements}. 

The functions $F_{M}$, $F_{\Sigma'}$, $F_{\Sigma''}$, $F_{\Delta}$, $F_{\tilde{\Phi}'}$ and $F_{\Phi''}$ are the standard nuclear form factors appearing in the study of semi-leptonic electroweak interactions \cite{Donnelly:1979} and are functions only of $q^2$ for a given nucleus. For the $^{19}F$ nucleus, we use the form factors given in Ref.~\cite{Fitzpatrick:2012ix} obtained under the one-body interaction approximation. Under this assumption, the form factors decay approximately exponentially with $q^2$, leading to suppression at large recoil energies. 

We assume isospin-zero interactions in this work, so only those form factors with $\tau = \tau'= 0$ will be relevant. In order to compare the relative strengths of the different form factors, we report below their values for $^{19}F$ at $q=0$:
\begin{align}
\begin{split}
F_M(0) &= 90.25 \\
F_{\Sigma'}(0) & = 0.435 \\
F_{\Sigma''}(0) &= 0.218 \\
F_{\tilde{\Phi}'}(0) &= 0 \\
F_{\Phi''}(0) &= 0.123 \\
F_{\Delta}(0) &= 0.0015\,.
\end{split}
\end{align} 
This means that where a response function contains contributions from two or more form factors, some may be subdominant. In the case of $\Op{8}$, for example, there are two terms in the recoil spectrum - one coupling to $v_\perp^2$ and the other to $q^2$. However, the form factors associated with each term ($F_M$ and $F_\Delta$) differ in normalisation by roughly 5 orders of magnitude. This means that the $q^2$-term in $F_{8,8}$ can effectively be neglected. 

\bibliographystyle{apsrev4-1}
\bibliography{DirectionalEFT.bib}

\end{document}